\documentstyle[preprint,aps,floats]{revtex}

\ifx\nopictures Y \else \input epsf.tex \fi
\input psfig 

\newcommand{\PLB}[3]{{\it Phys.~Lett.}~{\bf B#1},~#2~(19#3)}

\newcommand{\lesim}{\stackrel{\scriptscriptstyle<}{\scriptscriptstyle\sim}}

\def\beq{\begin{equation}}
\def\enq{\end{equation}}

\def\D0{D\O~}
\def\ETslash{\not{\hbox{\kern-4pt $E_T$}}}
\def\TblTotal{ 
\begin{table}
\begin{center}
\begin{tabular}{r r r r r r}
Di-boson   &
Collision   &
$\sqrt{S}$ &
\multicolumn{2}{c}{Fixed Order ${\cal O}(\alpha_s^0)$} &
Resummed \\
produced &
    type &
   (TeV) &
  CTEQ4L &
  CTEQ4M &
$\oplus$ ${\cal O}(\alpha _S)$
\\ \hline
$Z^0 Z^0$       & $      pp$ &  14 & 9.14 & 10.3 & 14.8 \\
$Z^0 Z^0$       & $p\bar{p}$ &   2 & 0.91 & 1.01 & 1.64 \\
$\gamma \gamma$ & $      pp$ &  14 & 22.1 & 24.5 & 60.8 \\
$\gamma \gamma$ & $p\bar{p}$ &   2 & 8.48 & 9.62 & 22.8 \\
$\gamma \gamma$ & $p\bar{p}$ & 1.8 & 6.30 & 7.15 & 17.0 \\
\end{tabular}
\end{center}
\par
\caption{Total cross sections of di-photon and $Z^0$ boson pair production
at the LHC and the upgraded Tevatron, in units of pb.
The kinematic cuts are described in the text.
The ``$\oplus$'' sign refers to the matching prescription discussed in the text.
}
\label{tbl:Total}
\end{table}
}
\def\TblSubTotalZZ{
\begin{table}
\begin{center}
\begin{tabular}{r r r r}
$\sqrt{S}$ & 
Collision &
$q\bar{q}\to  Z^0 Z^0 X$ &
$      qg\to  Z^0 Z^0 X$ 
\\ 
(TeV) &
type &    
 &
\\ \hline
 14 &       $pp$ & 10.9 & 3.91 \\ 
  2 & $p\bar{p}$ & 1.62 & 0.02 \\ 
\end{tabular}
\end{center}
\par
\caption{
Resummed cross sections of the subprocesses for $Z^0$ boson pair production
at the LHC and the upgraded Tevatron, in units of pb.
The kinematic cuts are described in the text.
}
\label{tbl:SubTotalZZ}
\end{table}
}
\def\TblSubTotalAA{
\begin{table}
\begin{center}
\begin{tabular}{r r r r r r r r}
$\sqrt{S}$ & 
Collision &
$q\bar{q}\to \gamma \gamma X$ &
$      qg\to \gamma \gamma X$ &         
 \multicolumn{2}{c}{$        gg\to \gamma \gamma$}  & 
$      gg\to \gamma \gamma g$ & 
$qg\to \gamma q X\to \gamma \gamma X'$ \\ 
(TeV) &
type &        
  &
  &
${\cal O}(\alpha_s^2)~4L$& 
${\cal O}(\alpha_s^2)~4M$& 
  &
Fragmentation \\
\hline
  14 &       $pp$ & 20.5 & 16.6 & 22.3 & 14.4 & 23.9 & 6.76 \\ 
   2 & $p\bar{p}$ & 9.68 & 4.81 & 6.02 & 4.34 & 8.26 & 2.15 \\ 
\end{tabular}
\end{center}
\par
\caption{
Cross sections of the subprocesses for di-photon production
at the LHC and the upgraded Tevatron, in units of pb.
The resummed $qg \to \gamma \gamma X$ rate includes
the fragmentation contribution.
The ${\cal O}(\alpha_s^2)$ $gg\to \gamma \gamma$ rates were calculated using
both the CTEQ4L and CTEQ4M PDF's.
The kinematic cuts are described in the text.
}
\label{tbl:SubTotalAA}
\end{table}
}
\def\FigDiagrams
{
\begin{figure*}[p]
\begin{center}
\begin{tabular}{c}
\ifx\nopictures Y \else{ \epsfysize=12.0cm \epsffile{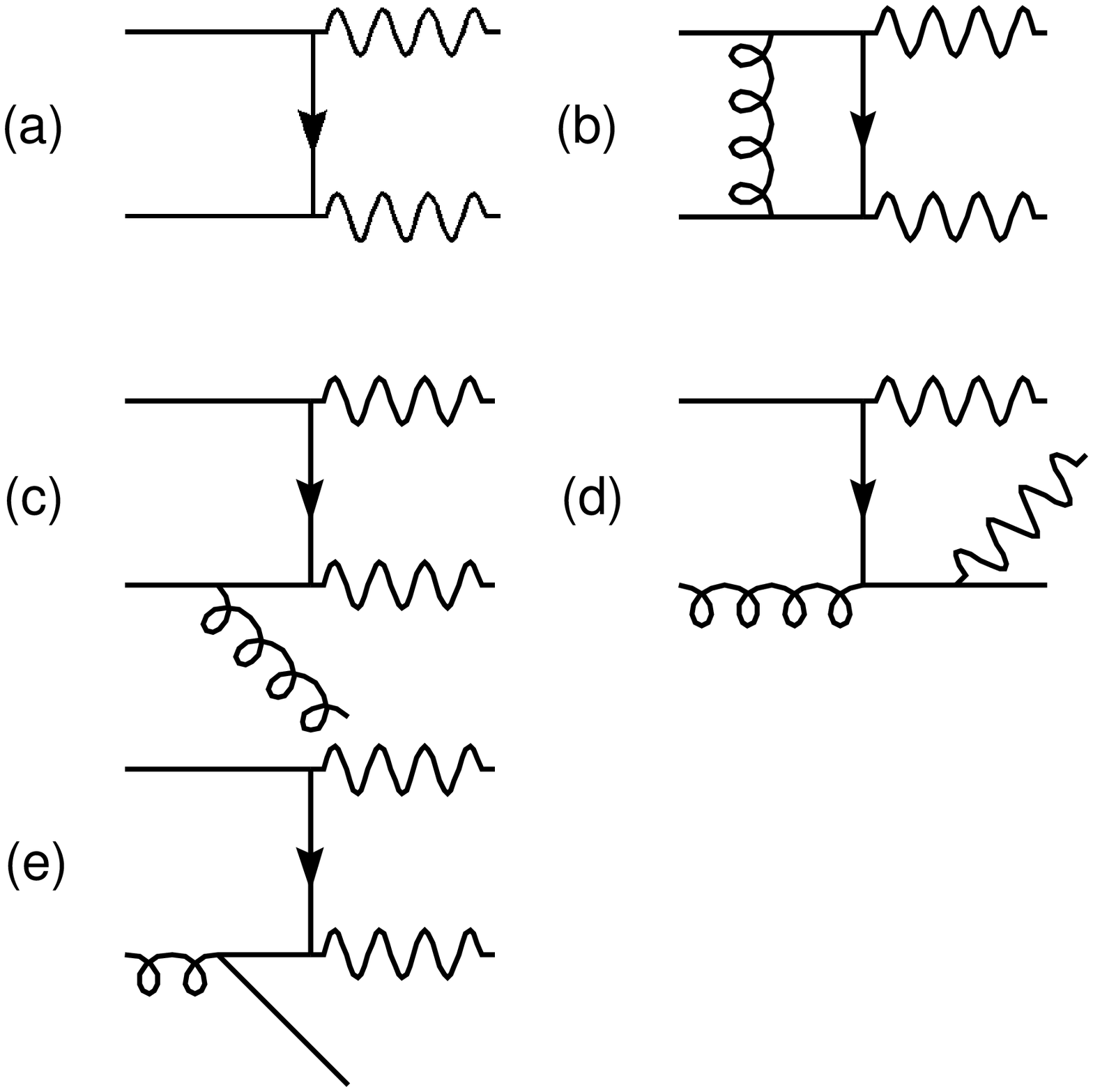}}
\fi 
\end{tabular}
\end{center}
\caption{ A representative set of Feynman diagrams included in the NLO
 calculation of $Z^0$ pair production. }
\label{fig:Diagrams}
\end{figure*}
}
\def\FigZZTevQypT
{
\begin{figure*}[p]
\begin{center}
\ifx\nopictures Y \else{
\begin{tabular}{cc}
\epsfysize=7.5cm \epsffile{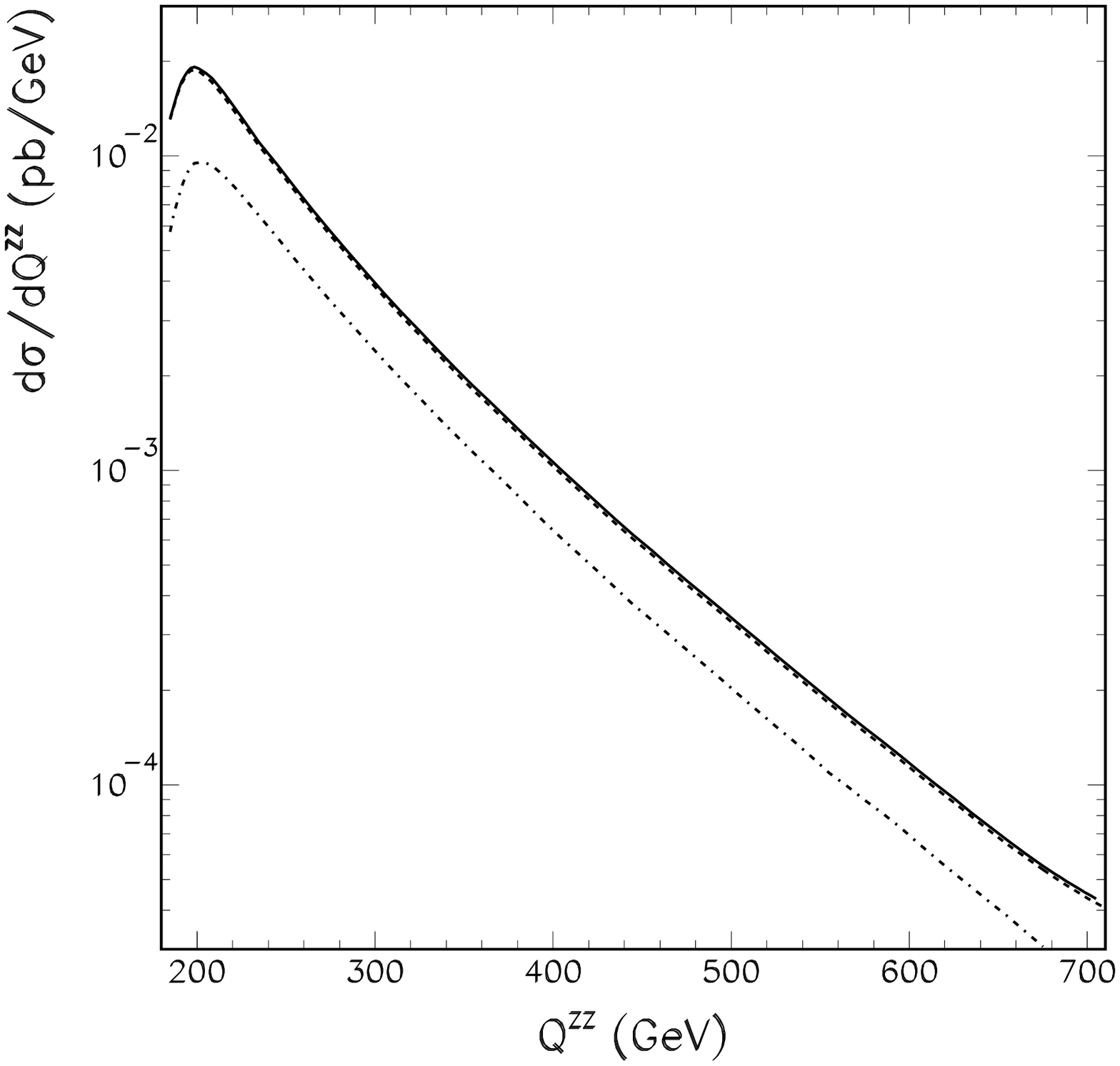} &
\epsfysize=7.5cm \epsffile{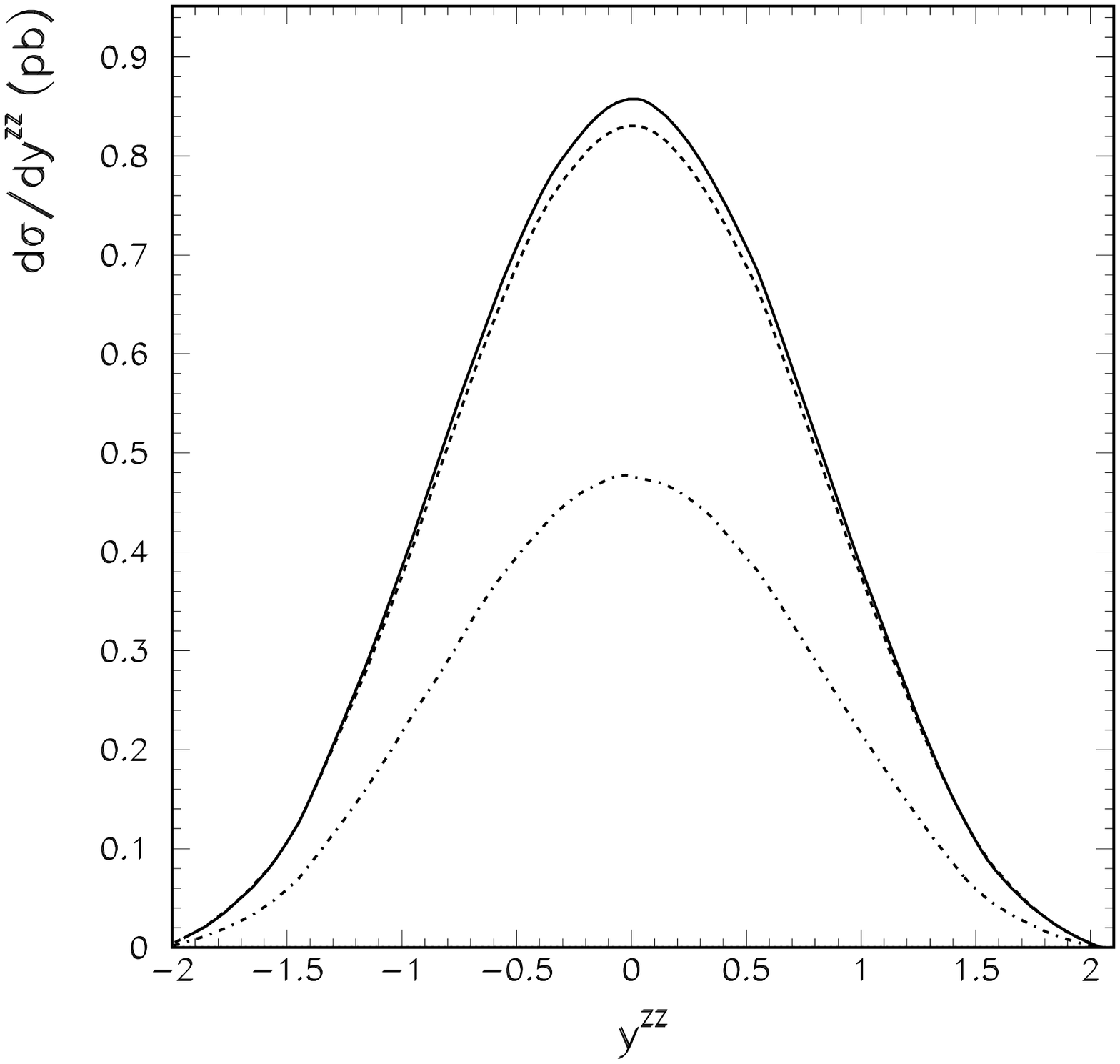} \\
\multicolumn{2}{c}{\epsfysize=7.5cm \epsffile{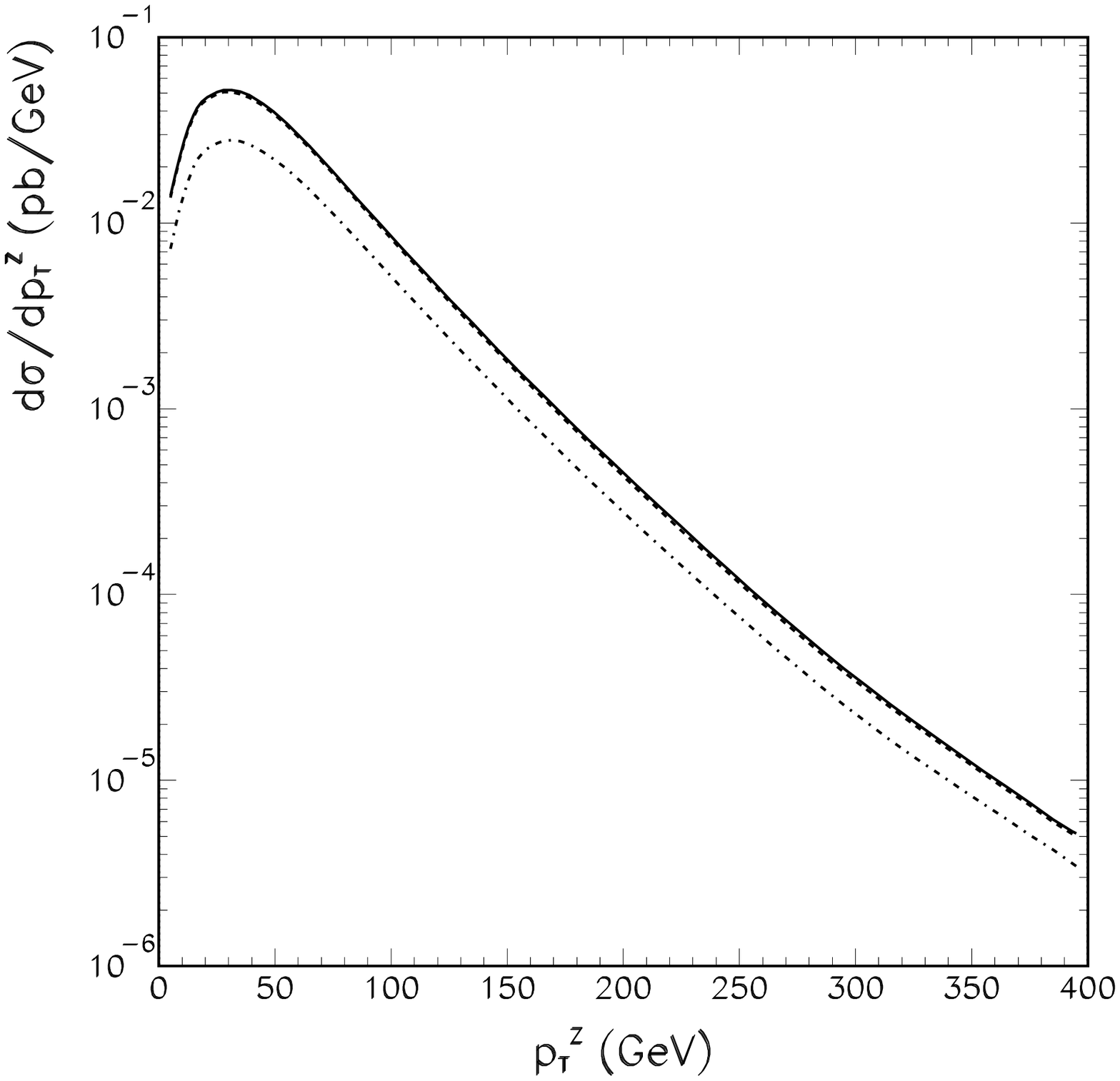} }
\end{tabular}
} \fi
\end{center}
\caption{
Same as Fig.~\ref{fig:ZZLHCQypT} except for the upgraded Tevatron.
}
\label{fig:ZZTevQypT}
\end{figure*}
}
\def\FigZZTevQT
{
\begin{figure*}[p]
\begin{center}
\begin{tabular}{c}
\ifx\nopictures Y \else{ \epsfysize=12.0cm \epsffile{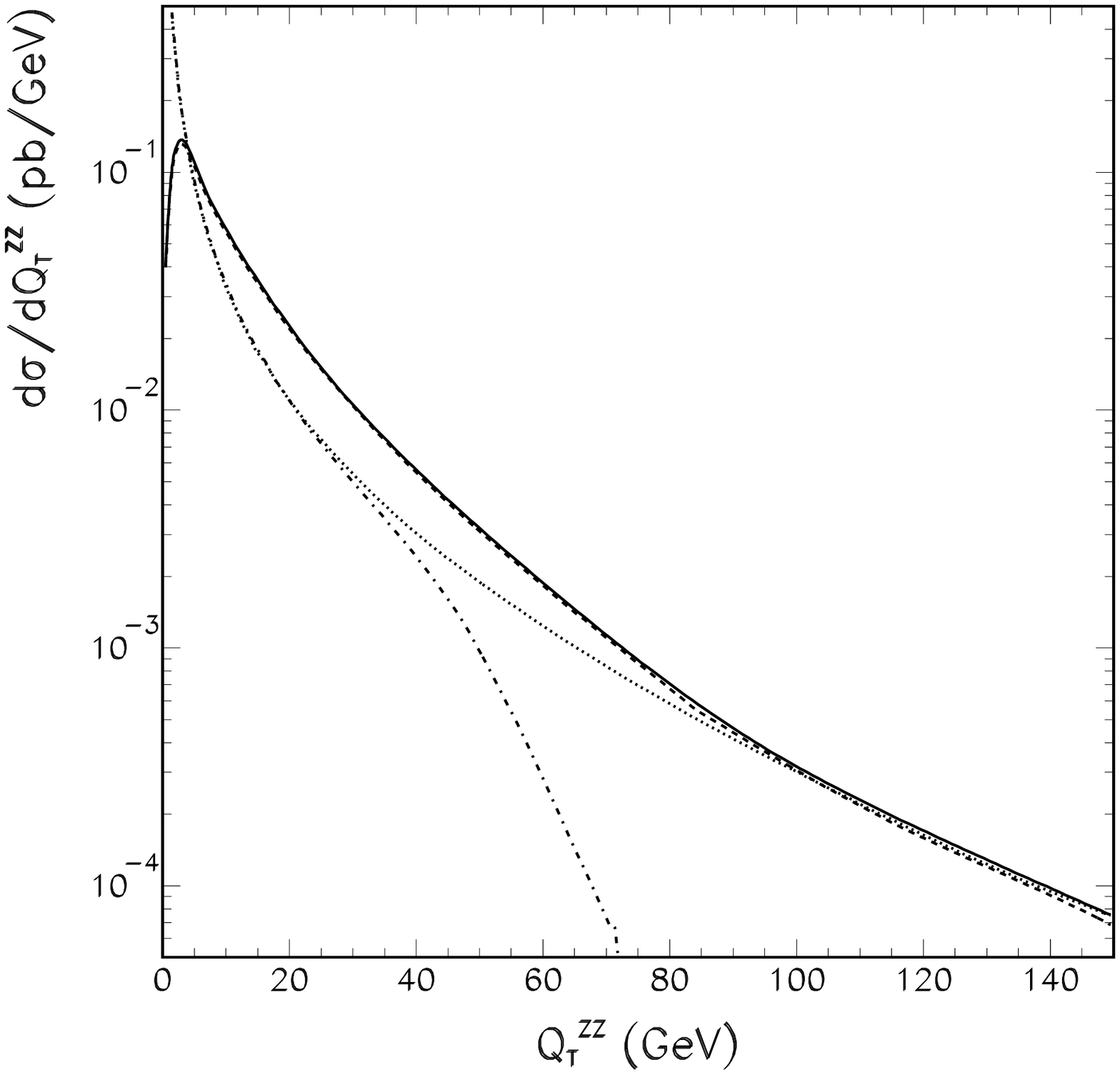}} 
\fi 
\end{tabular}
\end{center}
\caption{
Same as Fig.~\ref{fig:ZZLHCQT} except for the upgraded Tevatron.
}
\label{fig:ZZTevQT}
\end{figure*}
}
\def\FigZZTevInt
{
\begin{figure*}[p]
\begin{center}
\begin{tabular}{c}
\ifx\nopictures Y \else{ \epsfysize=12.0cm \epsffile{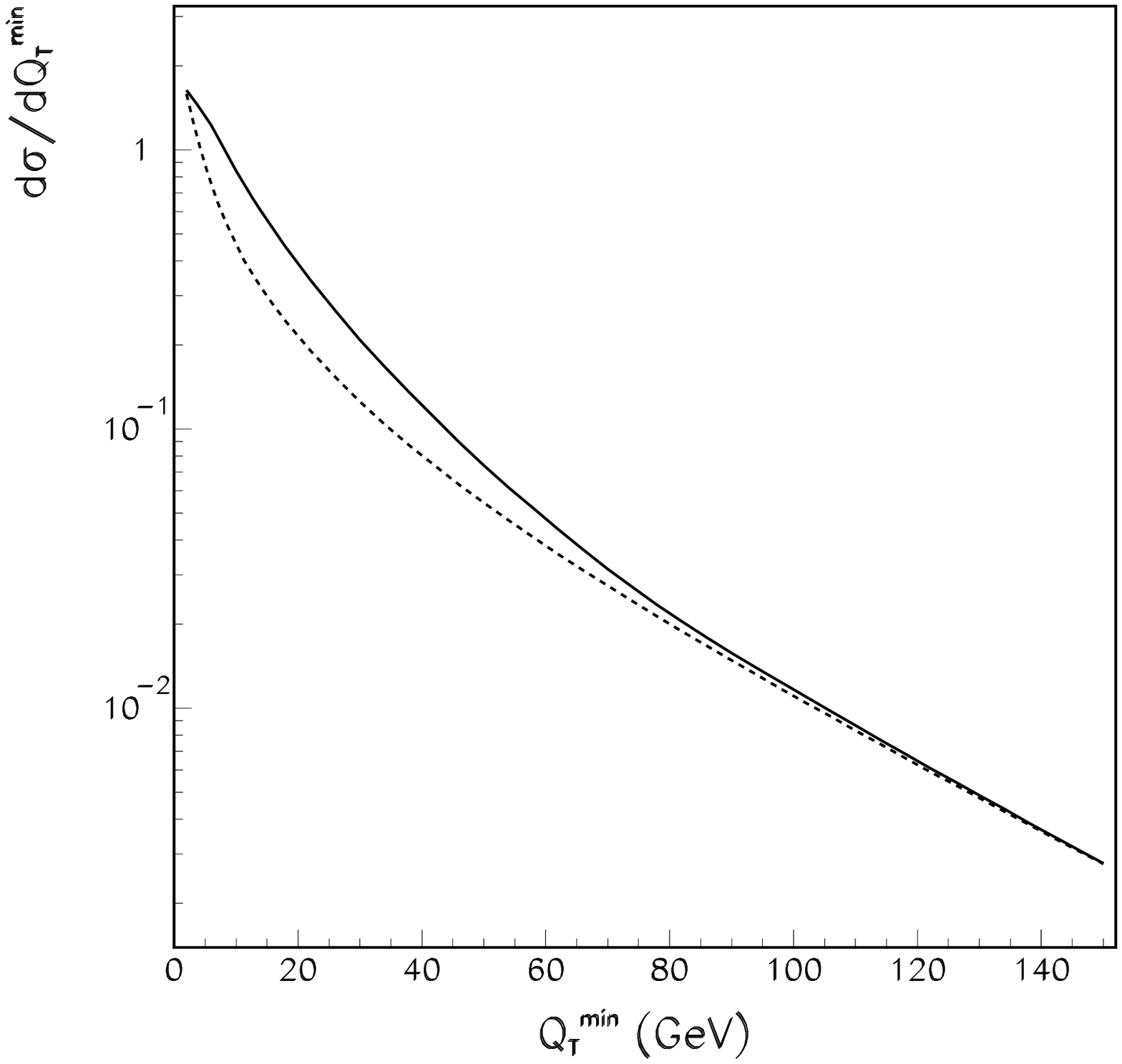}} \fi
\end{tabular}
\end{center}
\caption{
Same as Fig.~\ref{fig:ZZLHCInt} except for the upgraded Tevatron.
}
\label{fig:ZZTevInt}
\end{figure*}
}
\def\FigZZLHCQT
{
\begin{figure*}[p]
\begin{center}
\begin{tabular}{c}
\ifx\nopictures Y \else{ \epsfysize=12.0cm \epsffile{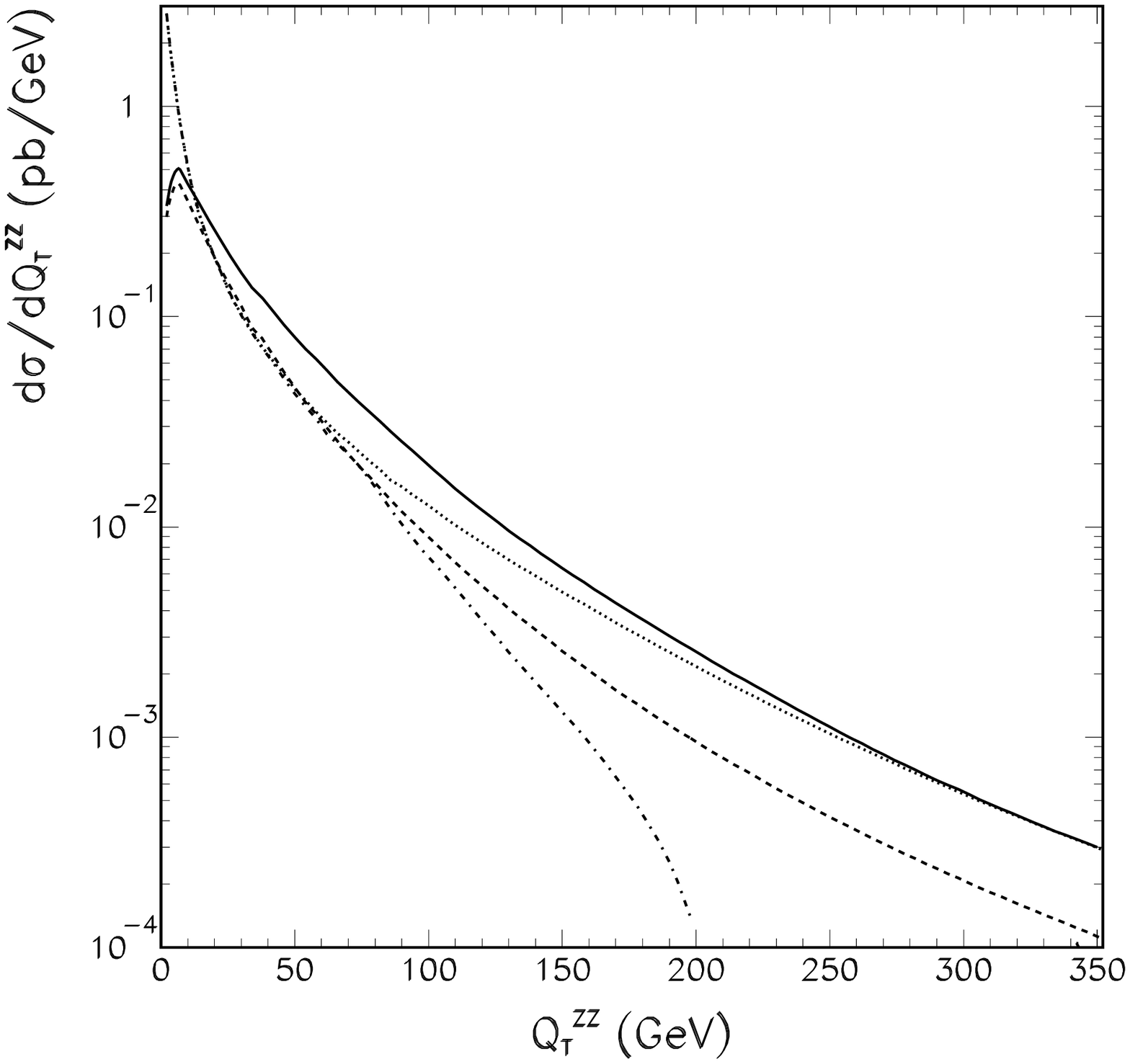}} \fi
\end{tabular}
\end{center}
\caption{
Transverse momentum distribution of $Z^0$ pairs from $q {\bar q} + q g$
partonic initial states at the LHC. The ${\cal O}(\alpha_s)$
(dotted) and the asymptotic (dash-dotted) pieces are coincide and
diverge as $Q_T \to 0$. 
The resummed (solid) curve matches the ${\cal O}(\alpha_s)$ curve at
about $Q_T = 320$ GeV. 
The resummed $q {\bar q}$ contribution (excluding the $qg$ contribution) 
is shown as dashed line.
}
\label{fig:ZZLHCQT}
\end{figure*}
}
\def\FigZZLHCInt
{
\begin{figure*}[p]
\begin{center}
\begin{tabular}{c}
\ifx\nopictures Y \else{ \epsfysize=12.0cm \epsffile{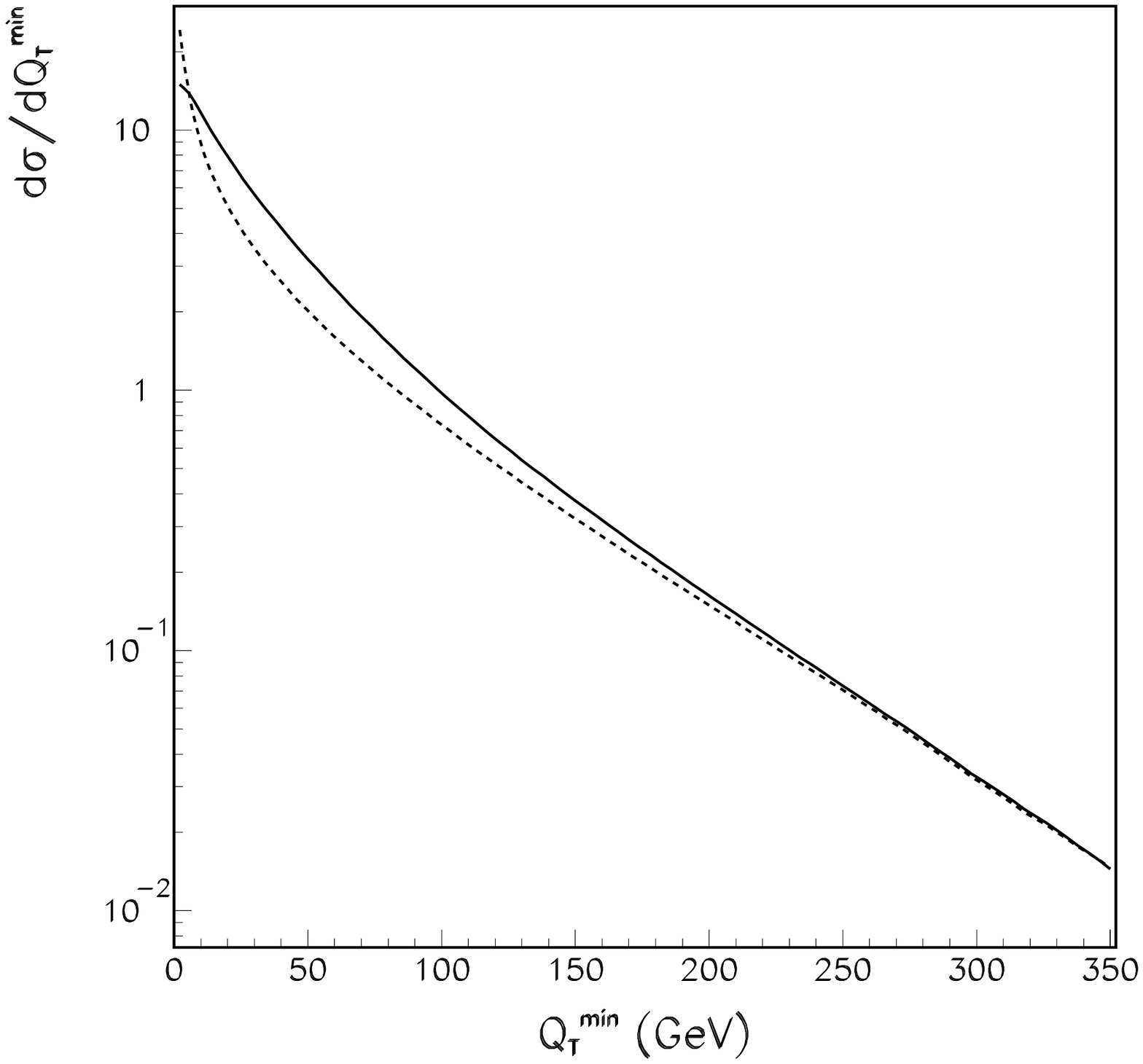}} \fi
\end{tabular}
\end{center}
\caption{
The integrated cross section for $Z^0$ boson pair production at the LHC. 
The resummed and the 
${\cal O}(\alpha_s)$ distributions are shown in solid and dashed lines, 
respectively.
 }
\label{fig:ZZLHCInt}
\end{figure*}
}
\def\FigZZLHCQypT
{
\begin{figure*}[p]
\begin{center}
\ifx\nopictures Y \else{
\begin{tabular}{cc}
\epsfysize=7.5cm \epsffile{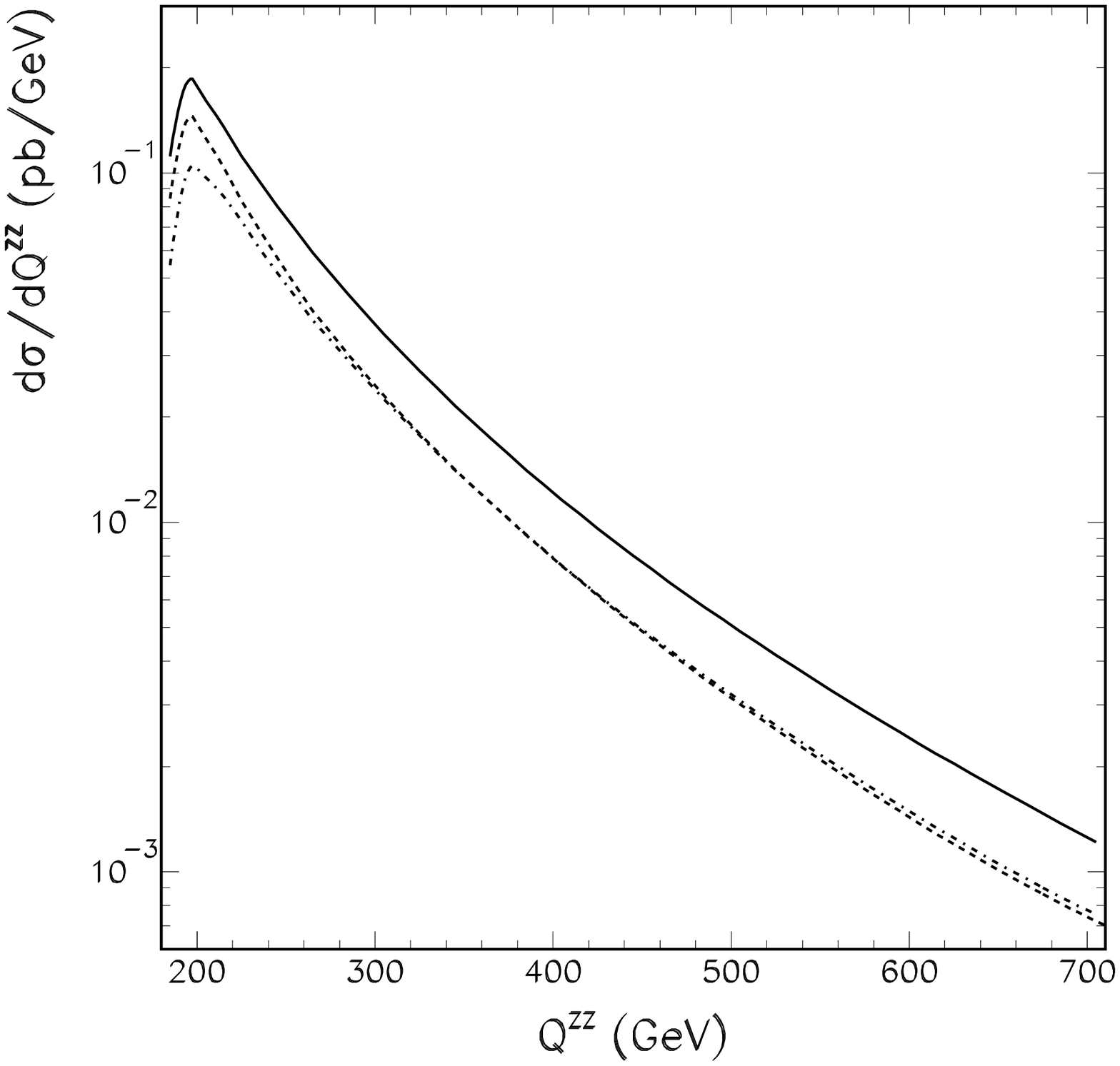} &
\epsfysize=7.5cm \epsffile{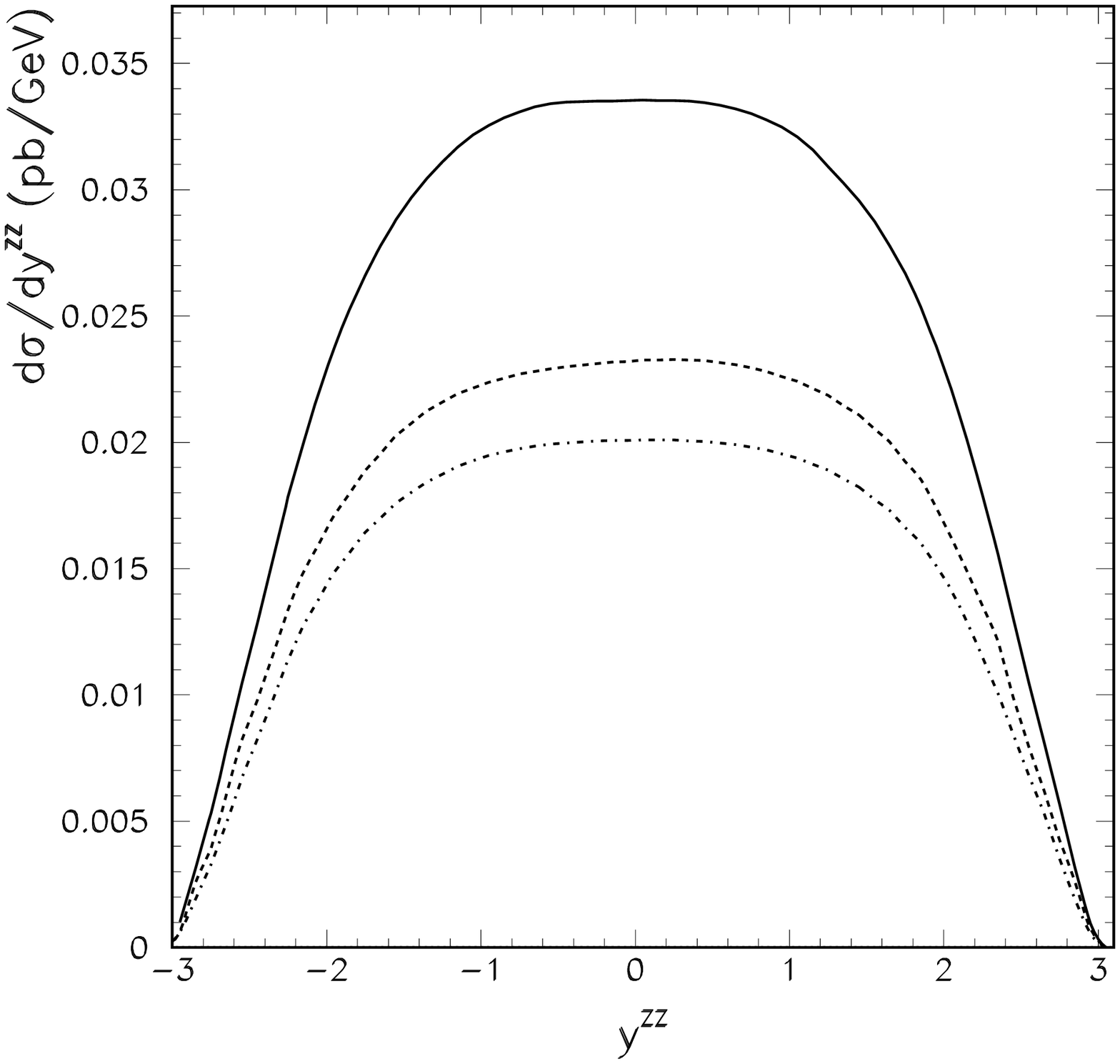} \\
\multicolumn{2}{c}{\epsfysize=7.5cm \epsffile{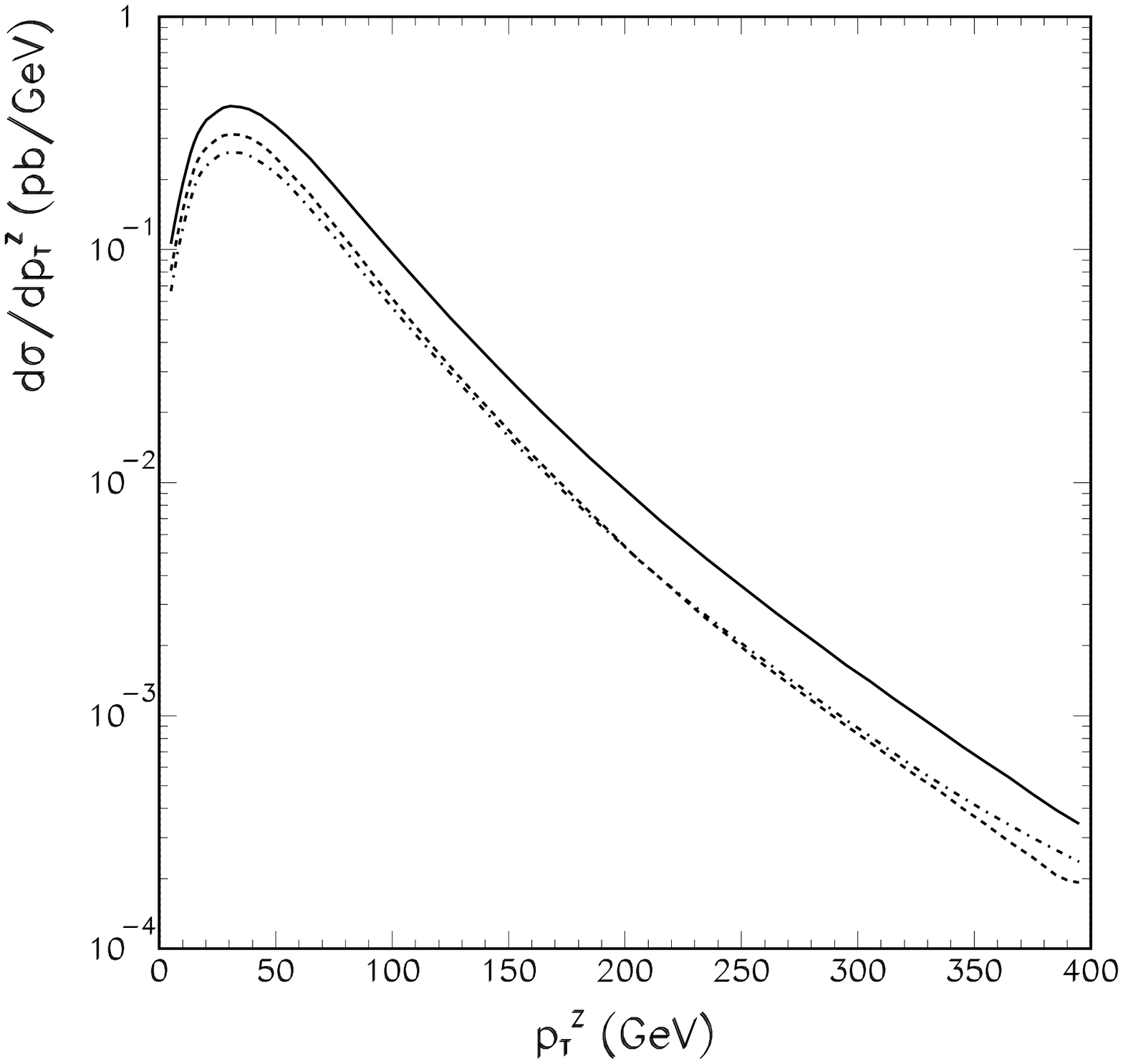} }
\end{tabular}
} \fi
\end{center}
\caption{
Invariant mass and rapidity distributions of $Z^0$ boson pairs, and 
transverse momentum distributions of the individual $Z^0$ bosons at the LHC. 
The resummed contribution of the $q {\bar q} + q g
\to Z^0 Z^0 X$ subprocess is shown by the solid curve, and of the 
$q \bar{q} \to Z^0 Z^0 X$ subprocess by the dashed curve. 
The leading order distribution of $q \bar{q} \to Z^0 Z^0$ is 
shown by the dash-dotted curve.
}
\label{fig:ZZLHCQypT}
\end{figure*}
}
\def\FigZZLHCQTCi
{
\begin{figure*}[p]
\begin{center}
\begin{tabular}{c}
\ifx\nopictures Y \else{ \epsfysize=12.0cm \epsffile{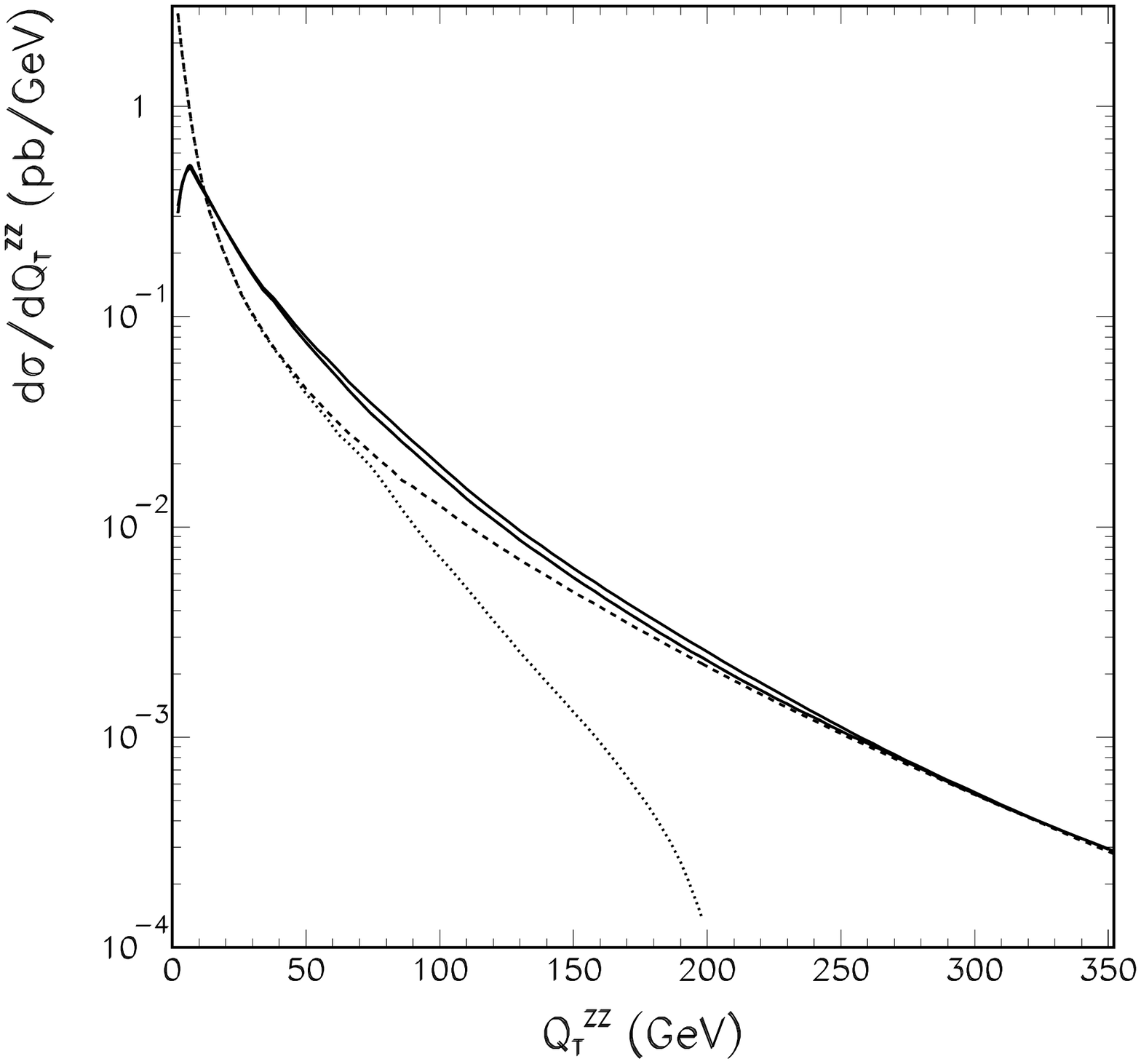}} \fi
\end{tabular}
\end{center}
\caption{
Resummed and NLO transverse momentum distributions of $Z^0$
boson pairs at the LHC. The two resummed curves are calculated for $C_1
= b_0$ and $C_2 = 1$ (upper solid), and for $C_1= b_0/2$ and $C_2 = 1/2$
(lower solid), respectively. The NLO curves are the same as in
Fig.~\ref{fig:ZZLHCQT}.
 }
\label{fig:ZZLHCQTCi}
\end{figure*}
}
\def\FigZZLHCQTgi
{
\begin{figure*}[p]
\begin{center}
\begin{tabular}{c}
\ifx\nopictures Y \else{ \epsfysize=12.0cm \epsffile{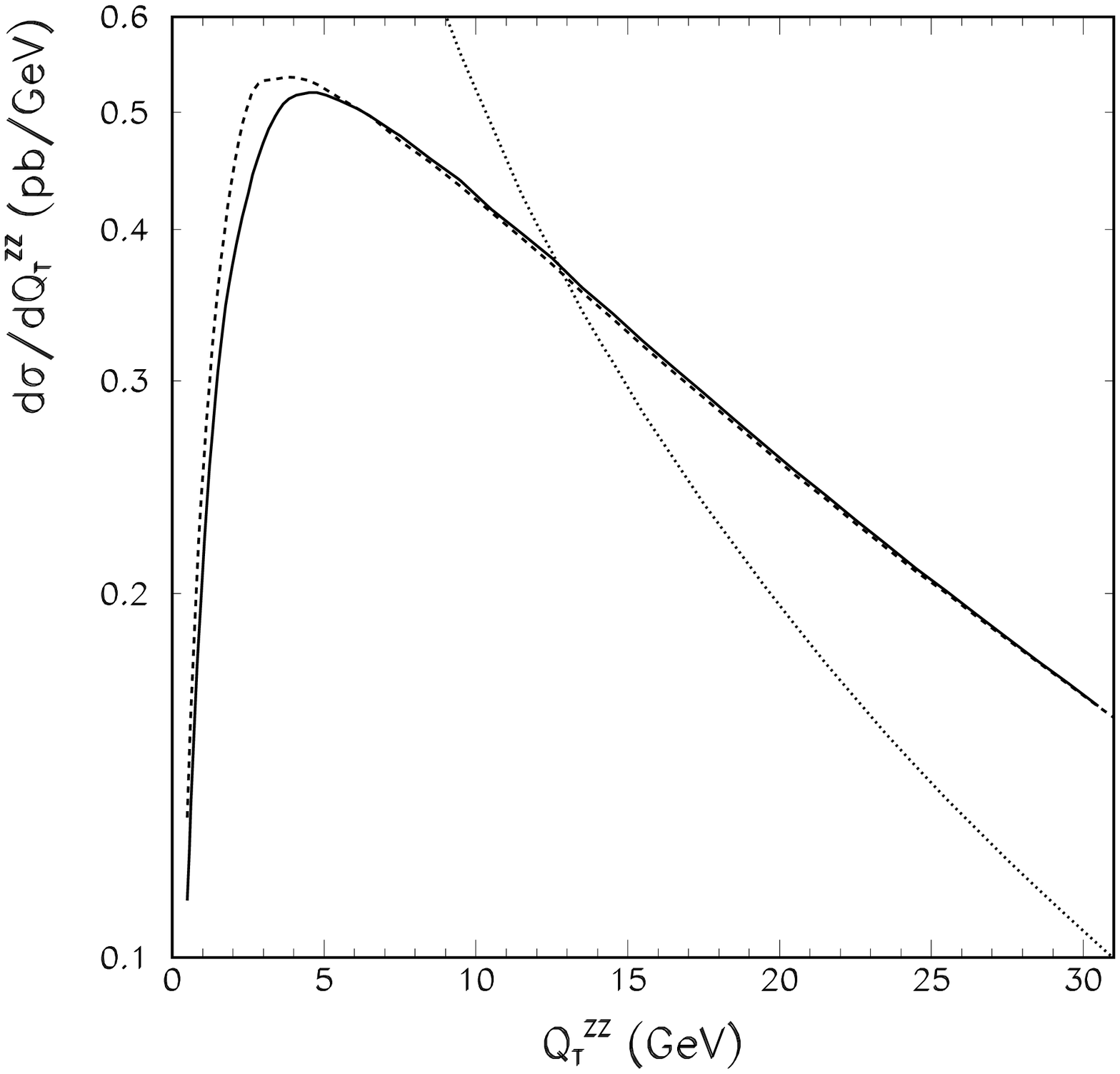}} \fi
\end{tabular}
\end{center}
\caption{
Resummed and NLO transverse momentum distributions of $Z^0$ boson pairs
at the LHC. The two resummed curves are calculated for $g_1=0.11~{\rm
GeV}^2$, $g_2=0.58~{\rm GeV}^2$, and $g_3=-1.5~{\rm GeV}^{- 1}$ (solid), and
for and for $g_1=0.15~{\rm GeV}^2$, $g_2=0.48~{\rm GeV}^2$, and
$g_3=-.58~{\rm GeV}^{-1}$ (dashed), respectively. In both cases $Q_0 =
1.6$ GeV was used. The NLO (dotted) curve is the same as in
Fig.~\ref{fig:ZZLHCQT}.
 }
\label{fig:ZZLHCQTgi}
\end{figure*}
}
\def\FigAATevQypT
{
\begin{figure*}[p]
\begin{center}
\ifx\nopictures Y \else{
\begin{tabular}{cc}
\epsfysize=7.5cm \epsffile{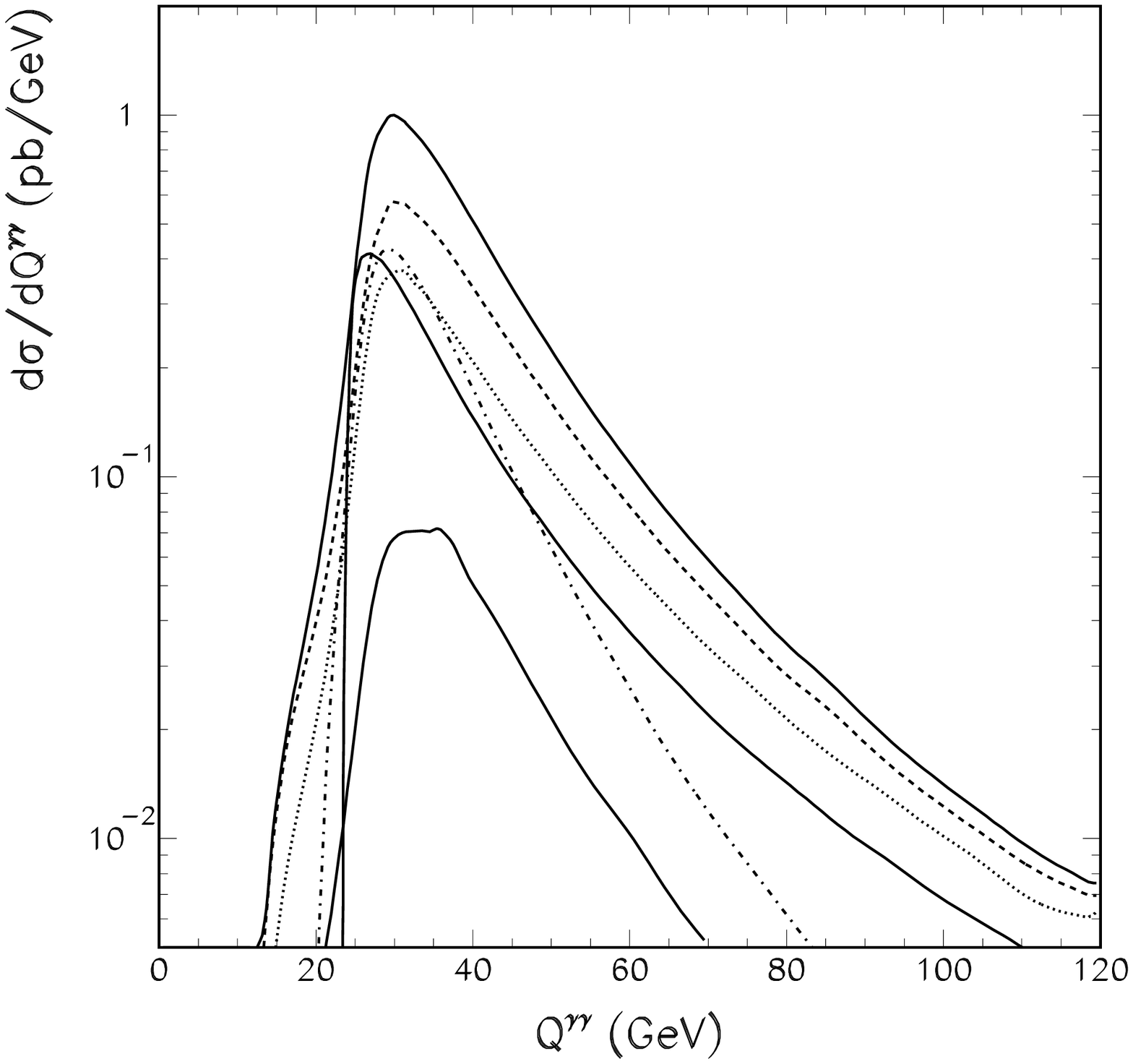} &
\epsfysize=7.5cm \epsffile{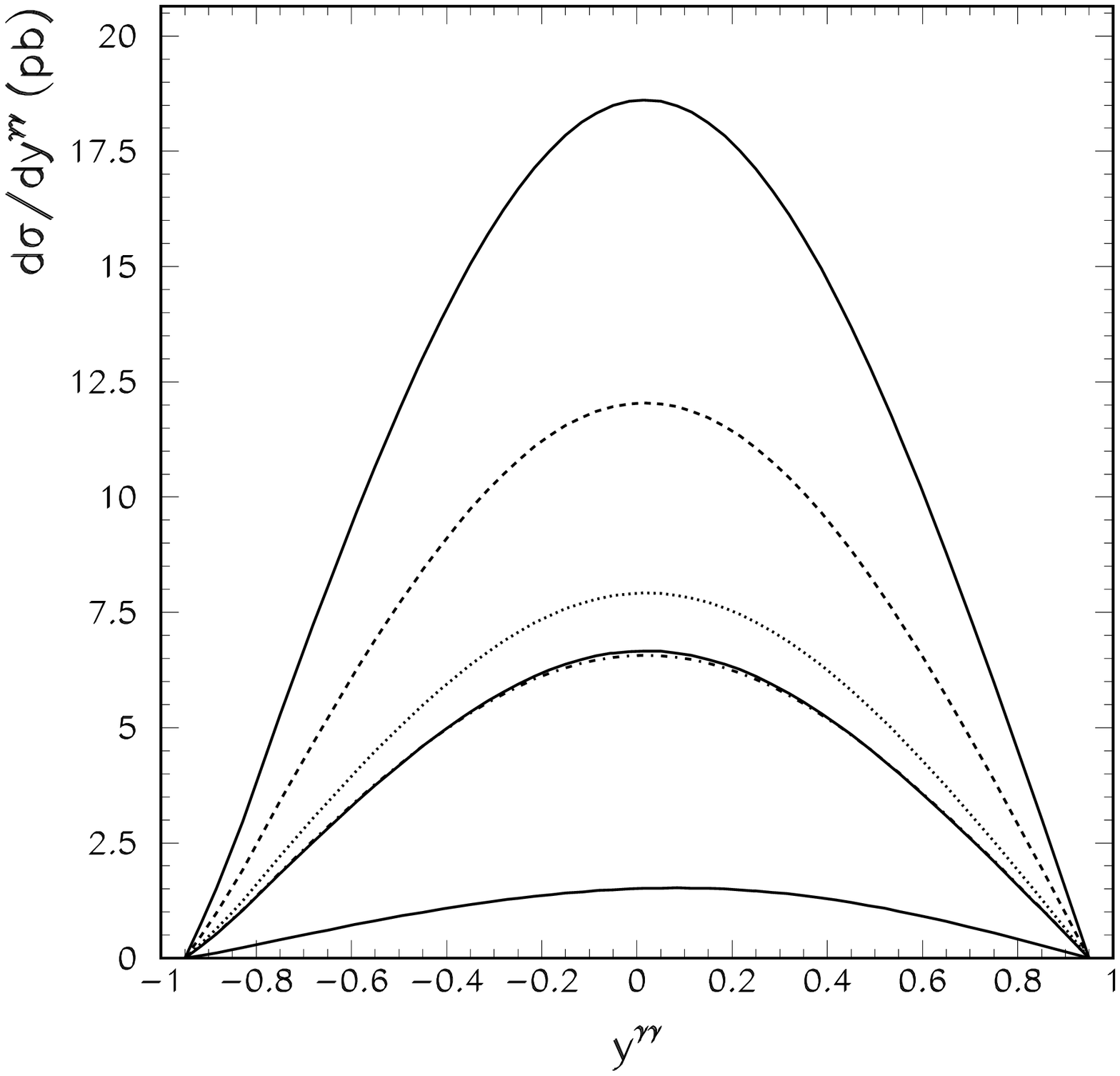} \\
\multicolumn{2}{c}{\epsfysize=7.5cm \epsffile{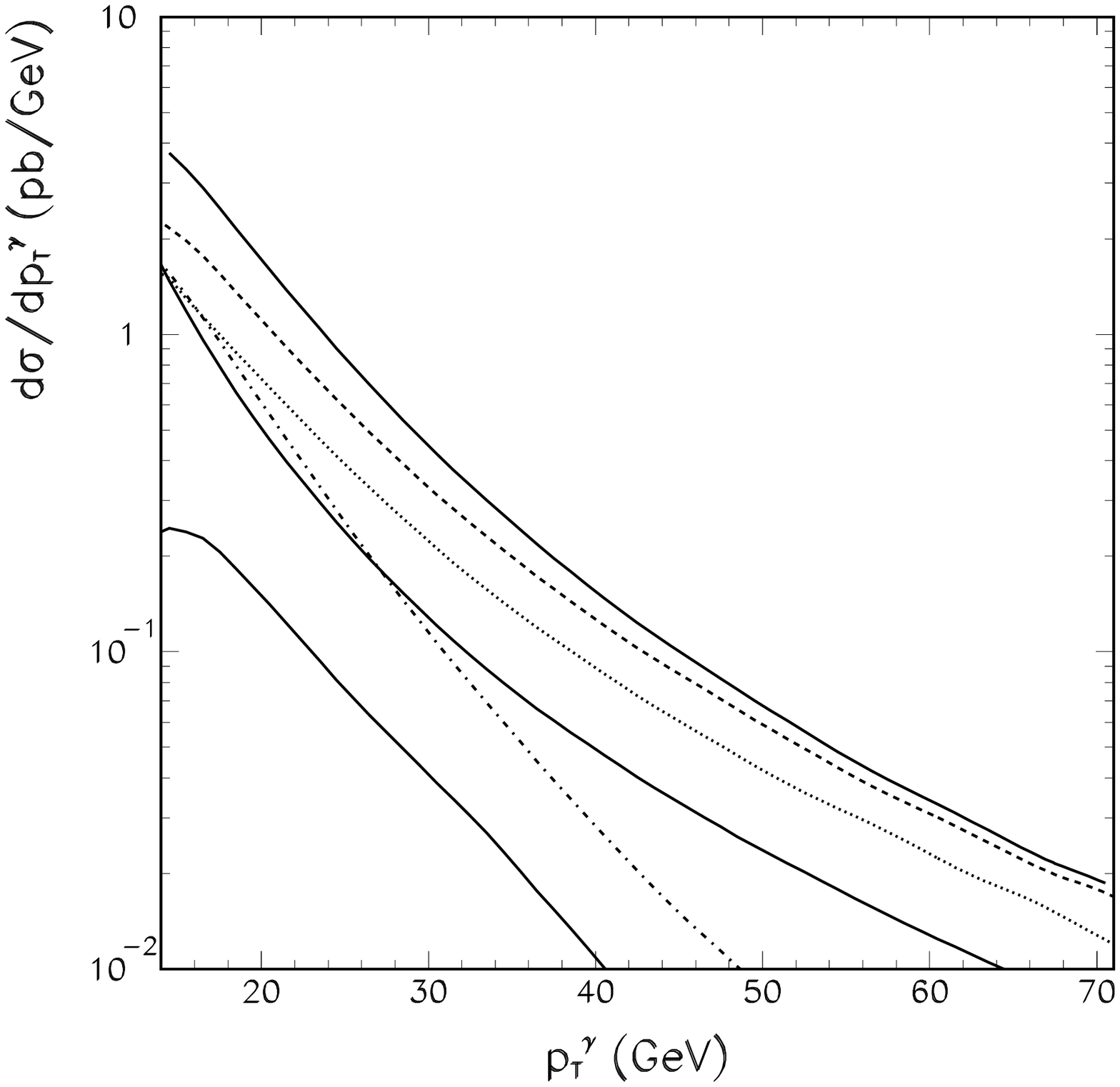} }
\end{tabular}
} \fi
\end{center}
\caption{
Same as Fig.~\ref{fig:AALHCQypT} but for the upgraded Tevatron.
}
\label{fig:AATevQypT}
\end{figure*}
}
\def\FigAATevQT
{
\begin{figure*}[p]
\begin{center}
\begin{tabular}{cc}
\ifx\nopictures Y \else{ \epsfysize=12.0cm \epsffile{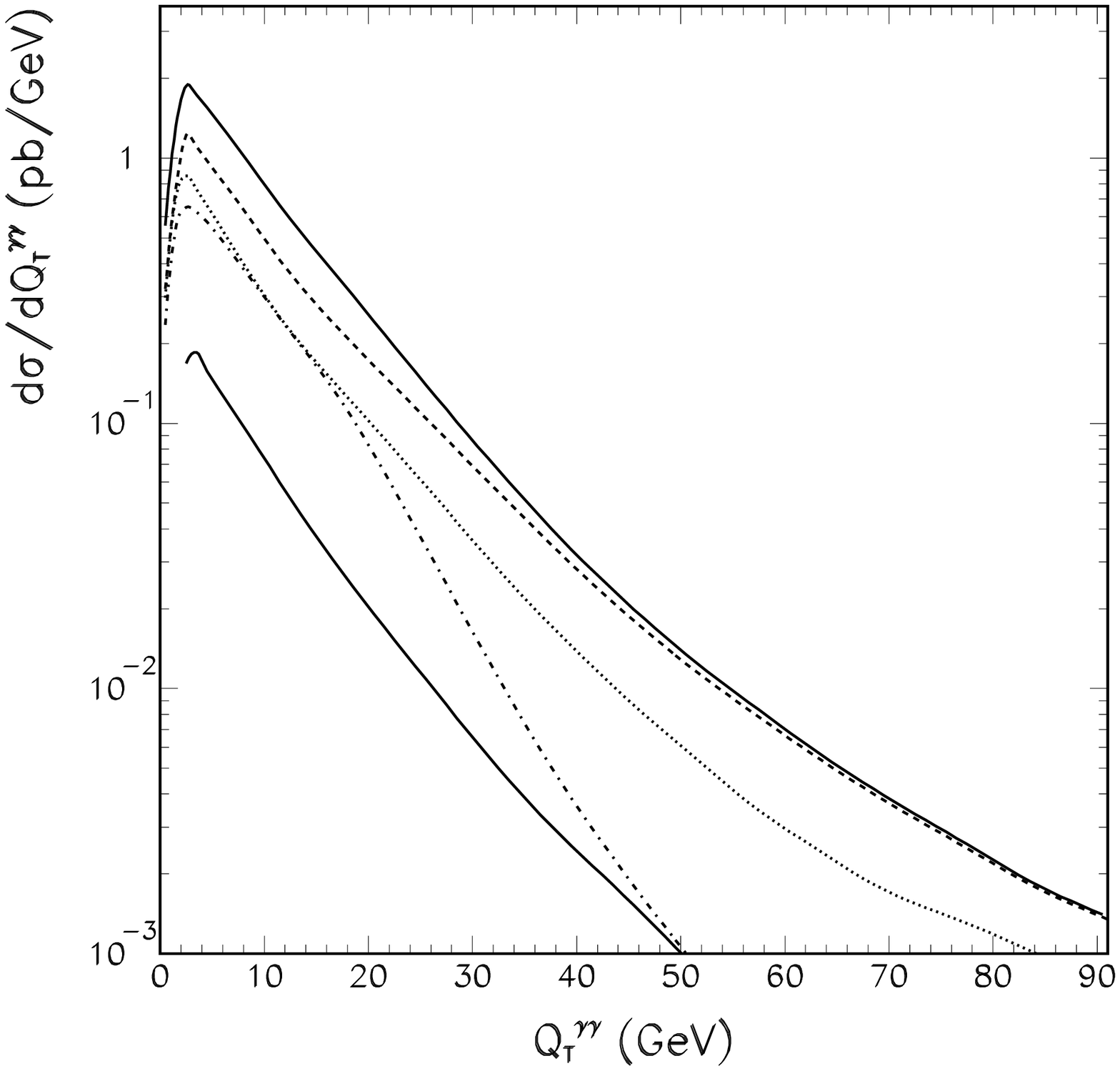}}
\fi & 
\end{tabular}
\end{center}
\caption{
Same as Fig.~\ref{fig:AALHCQT} but for the upgraded Tevatron.
}
\label{fig:AATevQT}
\end{figure*}
}
\def\FigAATevInt
{
\begin{figure*}[p]
\begin{center}
\begin{tabular}{c}
\ifx\nopictures Y \else{ \epsfysize=12.0cm \epsffile{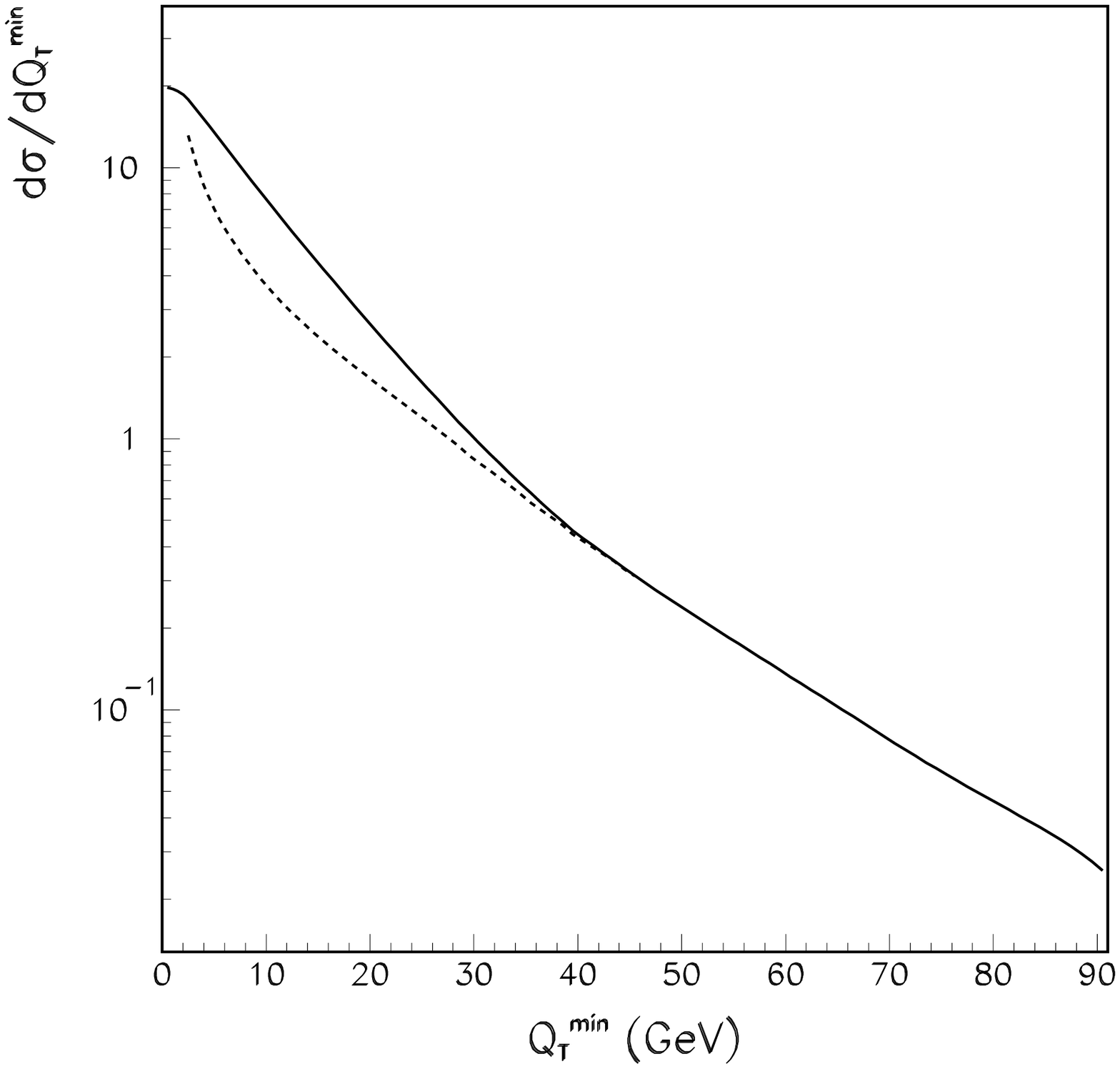}} \fi
\end{tabular}
\end{center}
\caption{
Same as Fig.~\ref{fig:AALHCInt} but for the upgraded Tevatron.
}
\label{fig:AATevInt}
\end{figure*}
}
\def\FigAALHCQypT
{
\begin{figure*}[p]
\begin{center}
\ifx\nopictures Y \else{
\begin{tabular}{cc}
\epsfysize=7.5cm \epsffile{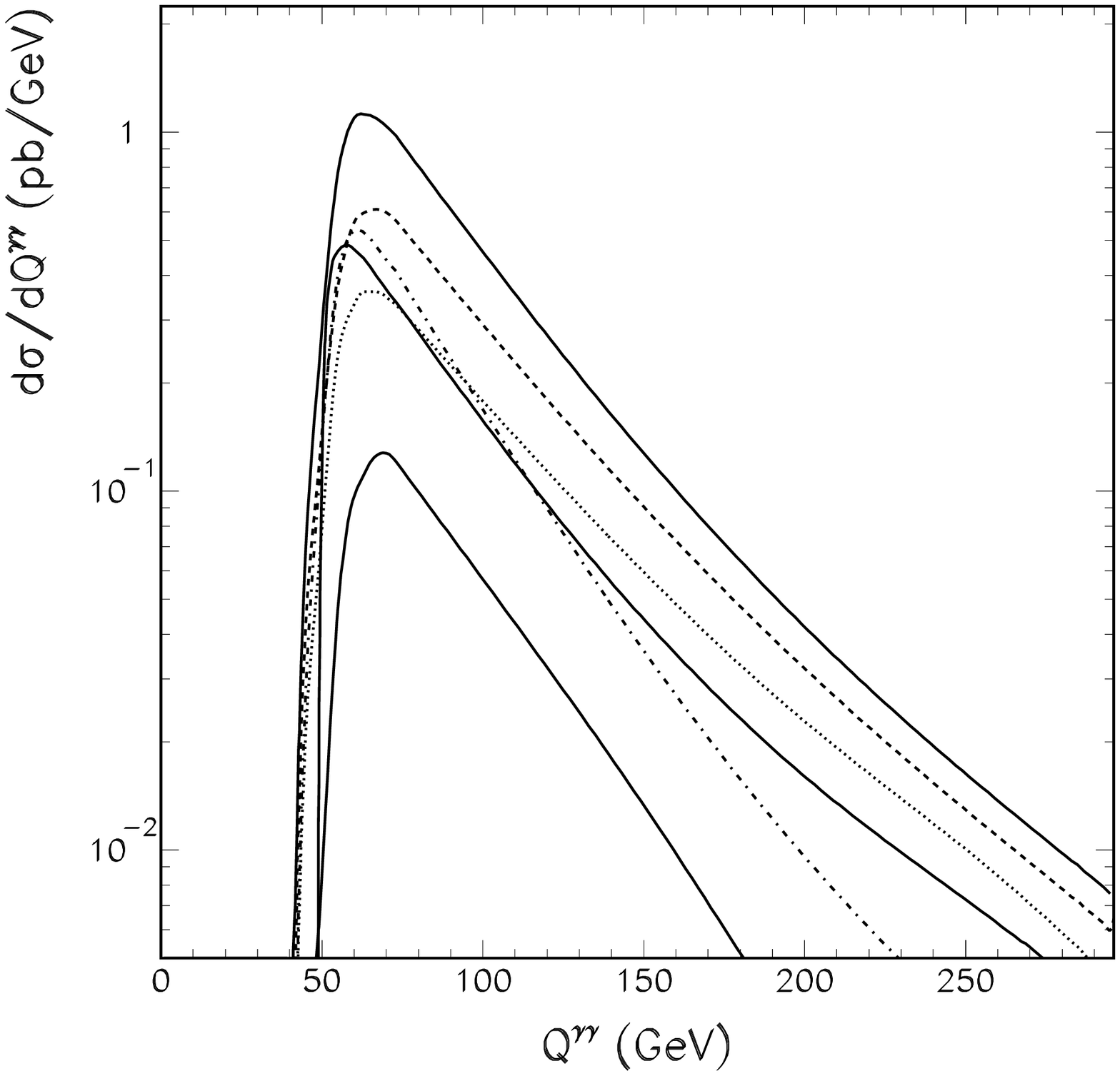} &
\epsfysize=7.5cm \epsffile{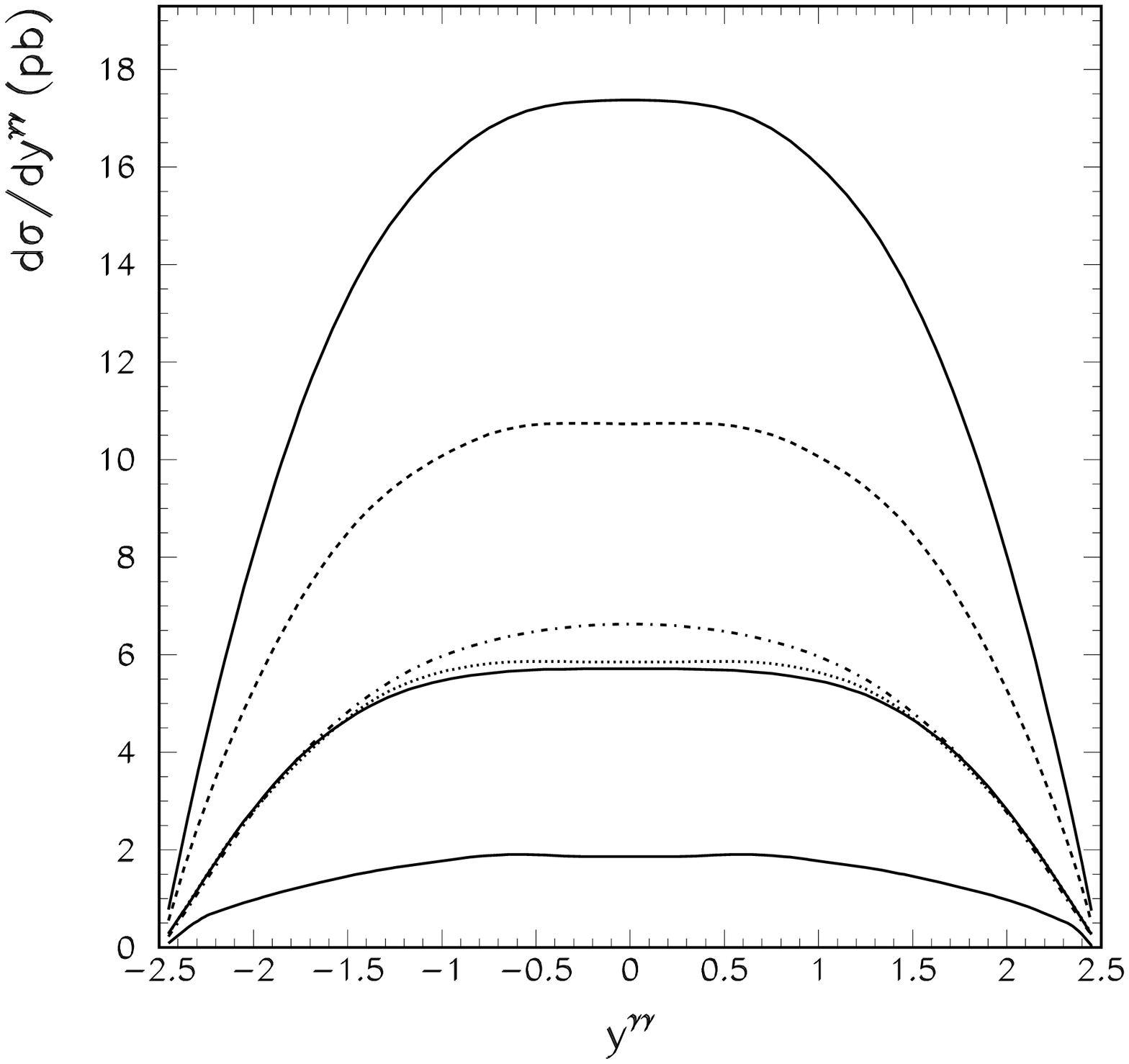} \\
\multicolumn{2}{c}{\epsfysize=7.5cm \epsffile{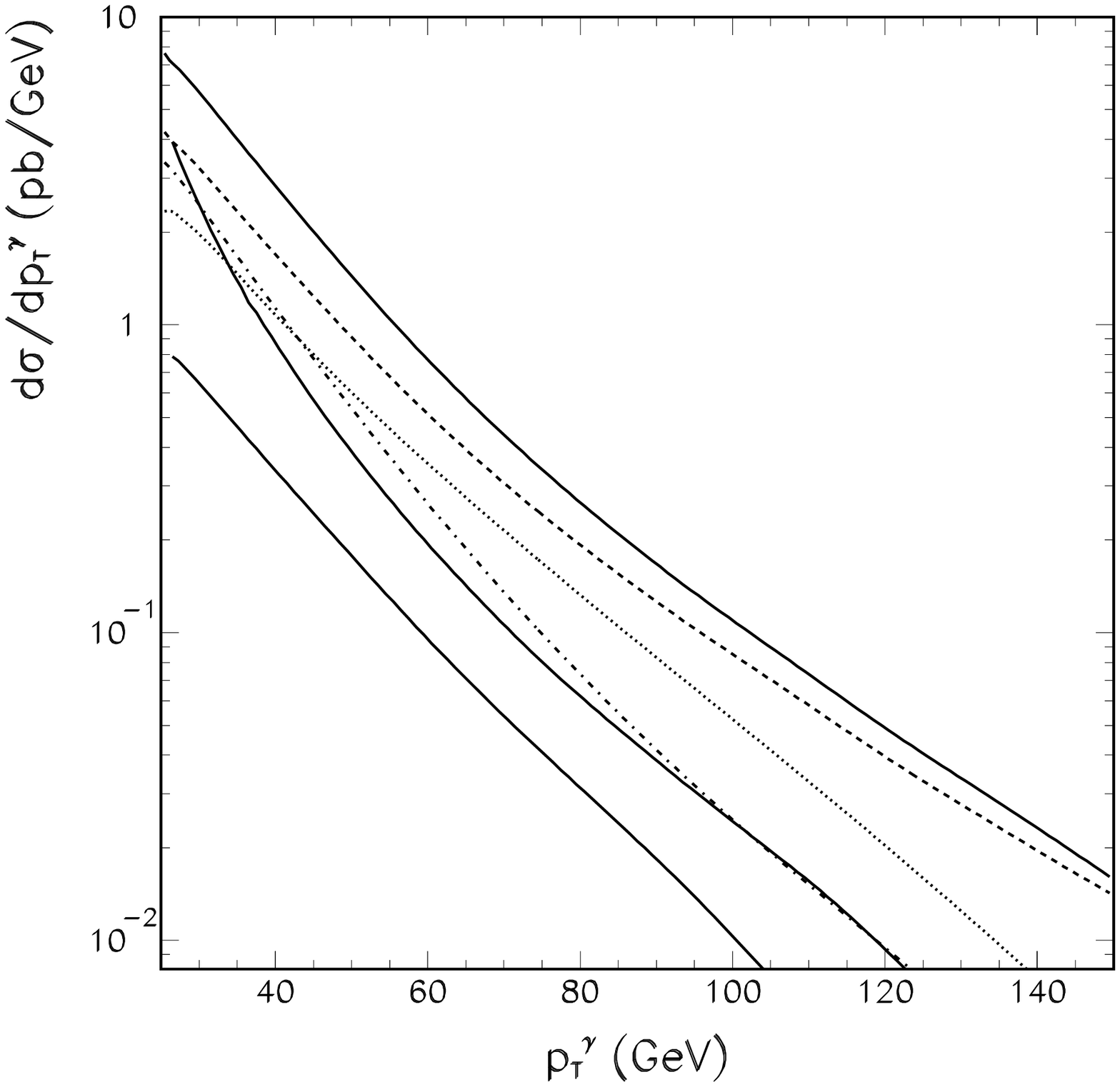} }
\end{tabular}
} \fi
\end{center}
\caption{
Invariant mass and rapidity distributions of photon pairs, and
transverse momentum distributions of the individual photons  
at the LHC. 
The total resummed contribution (upper solid), and the resummed 
$q{\bar q} + q g \to \gamma \gamma X$ (dashed), 
$q{\bar q}\to \gamma \gamma X$ (dotted), 
$g g \to \gamma \gamma g$ (dash-dotted), as well as the 
fragmentation (lower solid)
contributions are shown separately. 
The $q{\bar q}\to \gamma \gamma$ ${\cal O}(\alpha_s^0)$ 
distribution is shown in the middle solid curve. 
}
\label{fig:AALHCQypT}
\end{figure*}
}
\def\FigAALHCQT
{
\begin{figure*}[p]
\begin{center}
\begin{tabular}{c}
\ifx\nopictures Y \else{ \epsfysize=12.0cm \epsffile{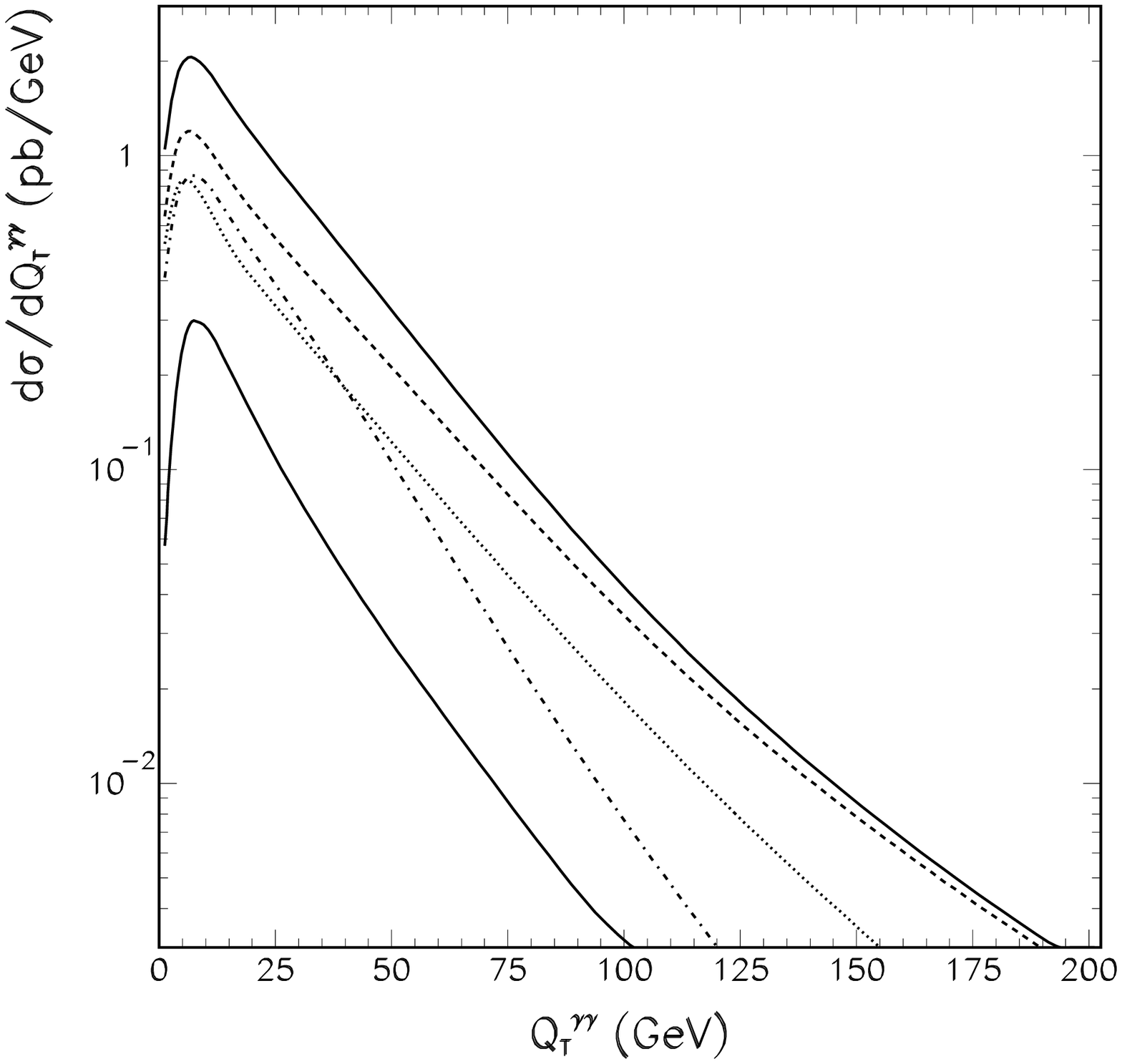}} \fi
\end{tabular}
\end{center}
\caption{Transverse momentum distribution of photon pairs at the LHC. 
The total resummed contribution (upper solid), the resummed
$q{\bar q} + q g \to \gamma \gamma X$ (dashed), 
$q{\bar q}\to \gamma \gamma X$ (dotted),
$g g \to \gamma \gamma X$ (dash-dotted), as well as the
fragmentation (lower solid) contributions are shown separately. }
\label{fig:AALHCQT}
\end{figure*}
}
\def\FigAALHCInt
{
\begin{figure*}[p]
\begin{center}
\begin{tabular}{c}
\ifx\nopictures Y \else{ \epsfysize=12.0cm \epsffile{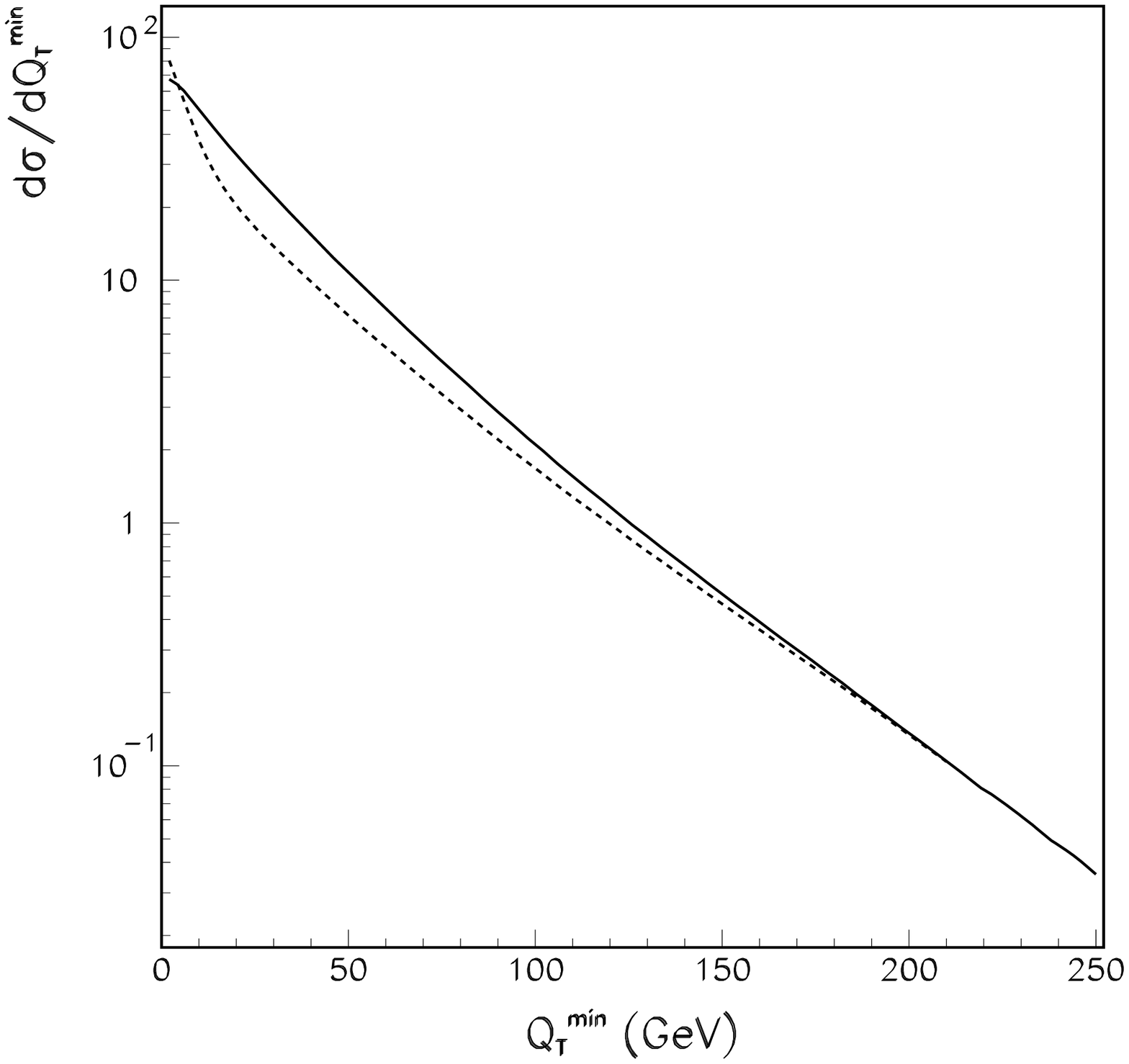}} \fi
\end{tabular}
\end{center}
\caption{The integrated cross section for photon pair production at the LHC. 
The resummed and the ${\cal O}(\alpha_s)$ distributions are shown in solid and
dashed lines, respectively.
 }
\label{fig:AALHCInt}
\end{figure*}
}

\begin{document}
\draft

\title{
Higgs Boson Production at Hadron Colliders \\
with Soft Gluon Effects: I. Backgrounds
}

\author{
C.~Bal\'azs$^{1,2}$\thanks{E-mail address: balazs@pa.msu.edu} and 
C.--P. Yuan$^{1}$\thanks{E-mail address: yuan@pa.msu.edu}}

\address{
$^1$~Michigan State University, East Lansing, MI 48824, U.S.A. \\
$^2$~Fermi National Accelerator Laboratory, Batavia, Illinois 60510, USA}
\maketitle

\thispagestyle{empty}

\begin{flushright}
{hep-ph/9810319  \hfill FERMILAB-PUB-98/202-T \\
                 \hfill CTEQ--815 \\
                 \hfill MSUHEP--80519}
\end{flushright}

\raggedbottom
\setcounter{page}{1}
\relax

\vskip -1.5cm

{\tighten
\begin{abstract}
The gold--plated discovery mode of a Standard Model like Higgs boson at
the CERN Large Hadron Collider (LHC) is the $H \to Z^0 Z^0$ decay mode.
To find and then measure the properties of the Higgs, it is crucial to
have the most precise theoretical prediction both for the signal and the
QCD background in this mode. In this work we calculate the effects of
the initial--state multiple soft--gluon emission on the kinematic
distributions of  $Z^0$ boson and photon pairs produced in hadron collisions.
The Collins--Soper--Sterman formalism is extended to resum the large
logarithmic terms due to soft--gluons. The resummed total rates, the
invariant mass, transverse momentum, and rapidity distributions of the $Z^0$
and photon pairs, and the transverse momentum distributions of the
individual vector bosons are presented and compared to the fixed order
predictions in the whole kinematic range, for the LHC energies and for 
the upgraded Fermilab Tevatron. Our conclusion is that the resummed
predictions should be used when extracting the Higgs signal at the LHC.
\end{abstract}}

\vspace{1cm}


\newpage

\section{Introduction}

The underlying dynamics of the electroweak symmetry breaking sector of
the Standard Model (SM) awaits understanding. The principal
goal of the CERN Large Hadron Collider (LHC) is to shed light on this open
question. 
The direct searches at the CERN Large Electron Positron (LEP) collider 
have constrained the mass of the SM Higgs boson to be higher than 
90 GeV \cite{Moriond98}. 
Furthermore, 
global analyses of electroweak data \cite{Langacker} and the
 values of the top
quark and the $W^{\pm }$ boson masses \cite{Parke} suggest 
that the SM Higgs boson is light, less than a few hundred GeV. 
Arguments based on supersymmetry (SUSY) also indicate that the
lightest Higgs boson is lighter than the top quark \cite{SUSYHiggsMass}.
Hence, the existence of a light Higgs boson is highly possible.
 
It has been shown in the literature that a SM like Higgs boson 
with a mass less than or about 120 GeV can be detected 
at the upgraded Fermilab Tevatron via 
$p\bar{p}\to W^{\pm} (\to \ell^{\pm} \nu)~H(\to b\bar{b},\tau^+ \tau^-)$
\cite{WHAtTev}, or
$p \bar p(gg) \to H(\to W^*W^* \to \ell \nu jj$ and $\ell \nu \ell \nu$)
\cite{Han-Zhang},
and a SUSY Higgs boson can be detected 
in the $W^{\pm} h$ and $hb\bar{b}$ modes 
\cite{Balazs-Diaz-He-Tait-Yuan}.
To observe a light ($m_H<120$ GeV) neutral Higgs
boson at the LHC, the most promising detection mode is the di-photon channel 
$H\to \gamma \gamma$ \cite{AtlasTechProposal}
via the production process $pp(gg)\to HX$.  
In the intermediate mass ($120\;{\rm GeV}<m_H<2~m_Z$)
region, the $Z^{0*}Z^0$ channel is also useful 
in addition to the $\gamma \gamma$ channel \cite{AtlasTechProposal}.
If the Higgs boson is heavier than twice the mass of the $Z^0$
boson, the gold--plated decay mode into two $Z^0$ bosons 
(which sequentially decay into leptons) \cite{AtlasTechProposal}
is the best way to detect it. 
At the LHC, as at any hadron--hadron collider, initial--state 
radiative corrections from
Quantum Chromodynamics (QCD) interaction to electroweak 
processes can be large. 
Some fixed order QCD corrections have been calculated
to the Higgs signal and to its most important backgrounds 
\cite{HiggsFixO,Ohnemus-Owens,Mele-Nason-Ridolfi,DiphotonFixO}. 
The next-to-leading order (NLO) corrections to the total cross section 
of $pp(gg)\to HX$ have been found to be large (50-100 \%) \cite{HiggsFixO}, 
and the largest contribution in the fixed order corrections results from 
soft gluon emission \cite{Kramer-Laenen-Spira}.
This signals the slow convergence of the perturbative series,
and the importance of still higher order corrections.
Furthermore, the
fixed order corrections fail to predict the transverse momentum
distributions of the
Higgs boson and its decay products correctly. The knowledge of these
distributions is necessary to precisely predict the signal and the
background in the presence of various kinematic cuts,
in order to deduce the accurate event rates to compare with theory 
predictions \cite{ADIKSS}. 
To predict the correct distribution of the transverse
momentum of the photon or $Z^0$ pair and the individual 
vector bosons, or the
kinematical correlation of the two vector bosons produced 
at hadron colliders, it is necessary to include
the effects from the initial--state multiple soft--gluon emission.
In this work, we present the results
of our calculation for the most important continuum backgrounds to the 
Higgs boson signal detected at hadron colliders. 
The distributions of the Higgs boson signal for the 
$h_1 h_2 \to H\to \gamma \gamma X$ and $Z^0 Z^0 X$ processes,
including the soft--gluon effects, 
will be discussed in our future work \cite{Balazs-Yuan_Higgs}.

In scattering processes involving hadrons, the dynamics of the multiple 
soft--gluon radiation can be described by the resummation technique.
We extend the Collins--Soper--Sterman (CSS) resummation formalism 
\cite{CS,CSS,Collins-Soper-Sterman} to describe the production of photon
and $Z^0$ pairs. 
This extension is analogous to our recent resummed calculation of
the hadronic production of photon pairs \cite{Balazs-Berger-Mrenna-Yuan}. 
In comparison, an 
earlier work \cite{Han-Meng-Ohnemus} on the soft--gluon resummation for the 
$q{\bar{q}}\to Z^0 Z^0 X$ process did not include the 
complete NLO corrections.
In the present work, the effect of 
initial--state multiple soft--gluon emission in 
$q{\bar{q}}\to Z^0 Z^0X$ is resummed with the inclusion of the full NLO 
contributions, so that the inclusive rate of the $Z^0$ boson pair 
production agrees with the NLO result presented 
in Ref.~\cite{Ohnemus-Owens}. 
Furthermore, we also include part of the higher order contributions 
in our results by using the CSS resummation formalism.

The collected di-photon data at the Tevatron, 
a $p\bar{p}$ collider with center-of-mass energy 
$\sqrt{S}=1.8$ TeV with 84 and 81 pb$^{-1}$ 
integrated luminosity (for CDF and \D0), 
are in the order of $10^3$ events per experiment 
\cite{CDFDiPhoton,D0DiPhoton}.
After the upgrade of the Tevatron, with $\sqrt{S}=2$ TeV
and a 2 fb$^{-1}$ integrated luminosity, about $4\times 10^4$ photon 
pairs can be detected, 
and more than $3\times 10^3$ $Z^0$ boson pairs can be produced. 
At the LHC, a $\sqrt{S}=14$ TeV $pp$ collider with 
a 100 fb$^{-1}$ integrated luminosity, 
we expect about $6\times 10^6$ photon 
and $1.5\times 10^6$ $Z^0$ pairs to be produced,
after imposing the kinematic cuts described later in the text.
This large data sample will play an important role
in the search for the Higgs boson(s) and new physics 
that modifies the production of the vector boson pairs
(e.g., by altering the vector boson tri-linear couplings
\cite{www}).

The rest of this paper is organized as follows. Section II briefly
summarizes the extension of the CSS resummation formalism to the $Z^0
Z^0$ pair production. In Section III, the numerical results of the
resummed and fixed order calculations are compared for various
distributions of the photon and $Z^0$ pairs produced at the LHC and the
upgraded Tevatron. Section IV contains our conclusions.

\section{Analytical Results}

\subsection{The CSS Resummation Formalism for $Z^0$ Pair Production}

When QCD corrections to the $Z^0$ boson pair production cross section are
calculated order by order in the strong coupling constant 
$\alpha _s$, 
the emission of potentially soft gluons spoils the convergence of the
perturbative series for small transverse momenta ($Q_T$) of the $Z^0$ boson
pair. In the $Q_T\ll Q$ region, the cross section can be written as \cite
{Collins-Soper-Sterman} 
\[
\lim_{{Q_T\to 0}}\frac{d\sigma }{dQ_T^2}=\sum_{n=1}^\infty
\sum_{m=0}^{2n-1}\alpha _s^n~\frac{_nv_m}{Q_T^2}~\ln ^m\left( 
\frac{Q^2}{Q_T^2}
\right) +{\cal O}\left (\frac 1{Q_T} \right), 
\]
where $Q$ is the invariant mass of the $Z^0$ boson pair, and 
the coefficients $_nv_m\,$are perturbatively calculable. 
At each order of the strong coupling the emitted
gluon(s) can be soft and/or collinear, which yields a small $Q_T$.
When the two scales $Q$ and $Q_T$ are very different, the
logarithmic terms $\ln ^m (Q^2/Q_T^2)$ are large, 
and for $Q_T\ll Q$ the perturbative series is
dominated by these terms. 
It was shown in Refs.~\cite{CS,CSS,Collins-Soper-Sterman} that 
these logarithmic contributions can be summed up to all order in $\alpha_s$, 
resulting in a well behaved cross section in the full $Q_T$ region.

The resummed differential cross section of the $Z^0$ boson pair production
in hadron collisions is written, similarly to the cross sections of the
lepton pair production \cite{Balazs-YuanPRD}, or photon pair production \cite
{Balazs-Berger-Mrenna-Yuan}, in the form: 
\begin{eqnarray}
&&{\frac{d\sigma (h_1h_2\to Z^0Z^0X)}{dQ^2\,dy\,dQ_T^2\,d\cos {\theta }%
\,d\phi }}={\frac 1{48\pi S}}\,{\frac {\beta}{Q^2}}  \nonumber \\
&&~~\times \left\{ {\frac 1{(2\pi )^2}}\int d^2b\,e^{i{\vec{Q}_T}\cdot
 {\vec{%
b}}}\,\sum_{i,j}{\widetilde{W}_{ij}(b_{*},Q,x_1,x_2,\theta ,\phi
,C_{1,2,3})}\,\widetilde{W}_{ij}^{NP}(b,Q,x_1,x_2)\right.  \nonumber \\
&&~~~~\left. +~Y(Q_T,Q,x_1,x_2,\theta ,\phi ,{C_4})\right\} .
\label{Eq:ResFor}
\end{eqnarray}
In this case, the variables $Q$, $y$, and $Q_T$ are the invariant mass,
rapidity, and transverse momentum of the $Z^0$ boson pair in the laboratory 
frame, while $\theta $ and $\phi $ are the polar and azimuthal angle
 of one of 
the $Z^0$ bosons in the Collins-Soper frame \cite{CSFrame}. The factor 
\[
\beta = \sqrt{1 -\frac{4 m_Z^2}{Q^2}} 
\]
originates from the phase space for producing the massive $Z^0$ boson pair. 
The parton
momentum fractions are defined as $x_1=e^yQ/\sqrt{S}$, and $x_2=e^{-y}Q/%
\sqrt{S}$, and $\sqrt{S}$ is the center--of--mass (CM) energy of the hadrons 
$h_1$ and $h_2$.

The renormalization group invariant function $\widetilde{W}_{ij}(b)$ sums
the large logarithmic terms $\alpha _s^n \ln ^m(b^2Q^2)$ to all orders in $%
\alpha _s$. For a scattering process initiated by the partons $i$ and $j$, 
\begin{eqnarray}
&&\widetilde{W}_{ij}(b,Q,x_1,x_2,\theta ,\phi ,C_{1,2,3})= \exp
\left\{ -{\cal S}_{ij}(b,Q,C_{1,2})\right\}  \nonumber \\
&& ~\times \left[ {\cal C}_{i/h_1}(x_1,b,C_{1,2,3},t,u)\, {\cal C}%
_{j/h_2}(x_2,b,C_{1,2,3},t,u) \right.  \nonumber \\
&& ~~ + \left. {\cal C}_{j/h_1}(x_1,b,C_{1,2,3},u,t)\, {\cal C}%
_{i/h_2}(x_2,b,C_{1,2,3},u,t)\right]  \nonumber \\
&& ~\times {\cal F}_{ij}(\alpha(C_2Q),\alpha _s(C_2Q),\theta
,\phi).  \label{Eq:WTwi}
\end{eqnarray}
Here the Sudakov exponent ${\cal S}_{ij}(b,Q,C_{1,2})$ is defined as 
\begin{equation}
{\cal S}_{ij}(b,Q,C_{1,2})=\int_{C_1^2/b^2}^{C_2^2Q^2}
{\frac{d{\bar{\mu}}^2}{%
{\bar{\mu}}^2}}\left[ A_{ij}\left( \alpha _s({\bar{\mu}}),C_1\right) \ln
\left( {\frac{C_2^2Q^2}{{\bar{\mu}}^2}}\right) +B_{ij}
\left( \alpha _s({\bar{%
\mu}}),C_1,C_2\right) \right] .  \label{Eq:SudExp}
\end{equation}
In Eq. (\ref{Eq:WTwi}), ${\cal F}_{ij}$ originates from the hard 
scattering process, and will be
given later for specific initial state partons.
 ${\cal C}_{i/h}(x)$ denotes
the convolution of the perturbative Wilson coefficient functions $%
C_{ia}$ with parton distribution functions (PDF) 
$f_{a/h}(\xi)$ (describing
the probability density of parton $a$ inside hadron $h$ 
with momentum fraction $\xi$): 
\begin{eqnarray}
&& {\cal C}_{i/h}(x,b,C_{1,2,3},t,u)=  \nonumber \\
&& \sum_a\int_{x}^1{\frac{d\xi}{\xi}}\,C_{ia} \left( {\frac{x}{\xi}}%
,b,\mu =\frac{C_3}b,C_1,C_2,t,u\right) \,f_{a/h}\left( \xi,\mu =\frac{C_3}%
b\right) .
\end{eqnarray}
The invariants $s$, $t$ and $u$ are defined for the $q(p_1)\bar{q}%
(p_2)\to Z^0(p_3)Z^0(p_4)$ subprocess as 
\begin{equation}
s=(p_1+p_2)^2,~~~~~~t=(p_1-p_3)^2,~~~~~~u=(p_2-p_3)^2,
\end{equation}
with $s+t+u=2m_Z^2$.

The functions $A_{ij}$, $B_{ij}$ and $C_{ij}$ are calculated
perturbatively in powers of $\alpha _s/\pi$:
\begin{eqnarray*}
A_{ij}\left( \alpha _s({\bar{\mu}}),C_1\right) &=&\sum_{n=1}^\infty \left( 
\frac{\alpha _s({\bar{\mu}})}\pi \right) ^nA_{ij}^{(n)}\left( C_1\right) , \\
B_{ij}\left( \alpha _s({\bar{\mu}}),C_1,C_2\right) &=&\sum_{n=1}^\infty
\left( \frac{\alpha _s({\bar{\mu}})}\pi \right) ^nB_{ij}^{(n)}\left(
C_1,C_2\right) , \\
C_{ij}\left( z,b,\mu ,C_1,C_2,t,u\right) &=& \sum_{n=0}^\infty \left( 
\frac{\alpha _s(\mu )}\pi \right) ^nC_{ij}^{(n)}\left(z,b,C_1,C_2,t,u\right) .
\end{eqnarray*}
The dimensionless constants $C_1,\;C_2$ and $C_3\equiv \mu b$ were
introduced in the solution of the renormalization group equations for $%
\widetilde{W}_{ij}$. Their canonical choice is $C_1=C_3=2e^{-\gamma
_E}\equiv b_0$, $C_2=C_1/b_0=1$, and $C_4=C_2=1$ 
\cite{Collins-Soper-Sterman}, where $\gamma _E$ is the Euler constant.

For large $b$, which is relevant for small $Q_T$, the 
perturbative evaluation 
of Eq.~(\ref{Eq:WTwi}) is questionable. 
Thus in Eq.~(\ref{Eq:ResFor}), $\widetilde{W}_{ij}$ is 
evaluated at $b_{*}={b/\sqrt{1+(b/b_{{\rm max}})^2}}$, 
so that the perturbative calculation is reliable.
Here $b_{{\rm max}}$ is a free parameter of the
formalism \cite{Collins-Soper-Sterman} 
that has to be constrained by other data (e.g. Drell--Yan),
along with the non-perturbative function $\widetilde{W}_{ij}^{NP}(b)$
which is introduced in Eq.~(\ref{Eq:ResFor}) to parametrize the 
incalculable long distance effects.
Since the $q{\bar q}\to \gamma \gamma$ or $Z^0 Z^0$, and the 
$q{\bar q}\to V \to \ell \ell '$ processes have the same initial state 
as well as the same QCD color structure, 
in this work we assume that the 
non-perturbative function $\widetilde{W}_{ij}^{NP}(b)$, fitted to existing 
low energy Drell-Yan data \cite{Ladinsky-Yuan}, 
also describes the non-perturbative 
effects in the $q{\bar q}\to \gamma \gamma$ and $Z^0 Z^0$ processes.
Needless to say, this assumption has to be tested by experimental data.

The function $Y$ in Eq.~(\ref{Eq:ResFor}) contains contributions from the
NLO calculation that are less singular than $1/Q_T^2$ or 
$\ln (Q^2/Q_T^2)/Q_T^2$ as $Q_T\to 0$. 
This function restores the regular contribution
in the fixed order perturbative calculation that is not included in the
resummed piece $\widetilde{W}_{ij}$. In the $Y$ function, both the
factorization and the renormalization scales are chosen to be $C_4Q$. The
detailed description of the matching (or ``switching'')
between the resummed and the fixed
order cross sections for $Q_T\sim Q$ can be found in
Ref.~\cite{Balazs-YuanPRD}.

\subsection{The $q\bar{q}$, $qg$ and $\bar{q}g\to Z^0Z^0X$ subprocesses}

The largest background to the Higgs boson signal 
in the $Z^0 Z^0$ channel is the continuum production
of $Z^0$ boson pairs via the $q{\bar{q}}\to Z^0Z^0X$ partonic subprocess 
\cite{Glover-Bij}. The next--to--leading order calculations of this process
are given in Refs. \cite{Ohnemus-Owens,Mele-Nason-Ridolfi}. 
A representative set of Feynman diagrams, 
included in the NLO calculations,
is shown in Fig.~\ref{fig:Diagrams}.
The application
of the CSS resummation formalism for the $q\bar{q}\to Z^0Z^0X $ subprocess
is the same as that for the case of $q\bar{q}\to \gamma \gamma X $ \cite
{Balazs-Berger-Mrenna-Yuan}. The $A^{(1)}$, $A^{(2)}$ and $B^{(1)}$
coefficients in the Sudakov exponent are identical to those of the
Drell--Yan case. This follows from the observation that to produce a heavy 
$Z^0$ boson pair, 
the virtual--quark line connecting the two $Z^0$ bosons in 
Fig.~\ref{fig:Diagrams} is far off the mass shell, and the leading logarithms
due to soft gluon emission beyond the leading order can only be generated
from the diagrams in which soft gluons are connected to the incoming
(anti--)quark. This situation was described in more detail for di-photon
production \cite{Balazs-Berger-Mrenna-Yuan}.

The resummed cross section is given by Eq.~(\ref{Eq:ResFor}), with 
$i$ and $%
j $ representing quark and anti--quark flavors, respectively, and 
\[
{\cal F}_{ij} (g,g_s,\theta,\phi) = 
2\delta _{ij}(g_L^2+g_R^2)^2 \, \frac{1+\cos ^2\theta}{1-\cos
^2\theta}. 
\]
The left- and right-handed couplings $g_{L,R}$ are defined through the $q{%
\bar{q}Z}^0$ vertex, which is written as $i\gamma _\mu \left[ g_L(1-\gamma
_5)+g_R(1+\gamma _5)\right]$, with 
\begin{equation}
g_L= g \, \frac{T_3-s_w^2Q_f}{2c_w} ~~~ {\rm and} ~~~ g_R=-g \, \frac{%
s_w^2Q_f}{2c_w}.  
\label{Eq:DefgLgR}
\end{equation}
Here $g$ is the weak coupling constant, $T_3$ is the third component of the
SU(2)$_L$ generator 
($T_3 = 1/2$ for the up quark $Q_u$, and $ -1/2$ for the down 
quark $Q_d$), 
$s_w$ ($c_w$) is the sine (cosine) of the weak mixing
angle, and $Q_f$ is the electric charge of the incoming quark in the units
of the positron charge ($Q_u=2/3$ and $Q_d=-1/3$). 
The values of these parameters will be given in the next section.

The explicit forms of the $A$
and $B$ functions, used in the numerical calculations are: 
\begin{eqnarray}
A_{q\bar{q}}^{(1)}(C_1) &=&C_F,  \nonumber \\
A_{q\bar{q}}^{(2)}(C_1) &=&C_F\left[ \left( \frac{67}{36}-\frac{\pi ^2}{12}%
\right) N_C-\frac 5{18}N_f-2\beta _1\ln \left( \frac{b_0}{C_1}\right)
\right] ,  \nonumber \\
B_{q\bar{q}}^{(1)}(C_1,C_2) &=&C_F\left[ -\frac 32-2\ln \left( \frac{C_2b_0}{%
C_1}\right) \right] ,
\label{Eq:Cjs}
\end{eqnarray}
where $N_f$ is the number of light quark flavors, $N_C=3$ is the number of
colors in QCD, $C_F=4/3$, and $\beta _1=(11N_C-2N_f)/12$.

To obtain the value of the total cross section to NLO, it is necessary to
include the Wilson coefficients $C_{ij}^{(0)}$ and $C_{ij}^{(1)}$, 
which can be
derived similarly to those for di-photon production \cite
{Balazs-Berger-Mrenna-Yuan}. The results are: 
\begin{eqnarray}
C_{jk}^{(0)}(z,b,\mu ,{\frac{C_1}{C_2}},t,u) &=&\delta _{jk}\delta ({1-z}),
\nonumber \\
C_{jG}^{(0)}(z,b,\mu ,{\frac{C_1}{C_2}},t,u) &=&0,  \nonumber \\
C_{jk}^{(1)}(z,b,\mu ,{\frac{C_1}{C_2}},t,u) &=&\delta _{jk}C_F\left\{
\frac 12(1-z)-\frac 1{C_F}\ln \left( \frac{\mu b}{b_0}\right) P_{j\leftarrow
k}^{(1)}(z)\right.  \nonumber \\
&&\left. +\delta (1-z)\left[ -\ln ^2\left( {\frac{C_1}{{b_0C_2}}}%
e^{-3/4}\right) +{\frac{{\cal V}(t,u)}4}+{\frac 9{16}}\right] \right\} , 
\nonumber \\
C_{jG}^{(1)}(z,b,\mu ,{\frac{C_1}{C_2}},t,u) &=&{\frac 12}z(1-z)-\ln
\left( \frac{\mu b}{b_0}\right) P_{j\leftarrow G}^{(1)}(z).
\label{eq:quarks}
\end{eqnarray}
In the above expressions, the splitting kernels are \cite{DGLAP}  
\begin{eqnarray}
P_{{j\leftarrow k}}^{(1)}(z) &=&C_F\left( {\frac{1+z^2}{{1-z}}}\right)_+ ~~~%
{\rm and}  \nonumber \\
P_{{j\leftarrow G}}^{(1)}(z) &=&{\frac 12}\left[ z^2+(1-z)^2\right] .
\label{eq:DGLAP}
\end{eqnarray}
For $Z^0$ boson pair production the function ${\cal V}$ in Eq.~(\ref{Eq:Cjs})
is given by 
\[
{\cal V}(t,u)={\cal V}_{Z^0Z^0}(t,u)=    \frac{\pi^2}{3}+|{\cal M}^{\rm Born}%
_0|^{-2} \left(- |{\cal M}^{\rm Born}_2|^2 + 4 {\cal N} (F(t,u)+F(u,t)) \right) 
- 3 .
\]
The squares of the invariant amplitudes ${\cal M}^{\rm Born}_0$ and 
${\cal M}^{\rm Born}_2$ are defined by Eqs.(6) and (7) of 
Ref.~\cite{Ohnemus:1991za}. 
For $Z^0$ pair production 
${\cal N} = N_C Q_{e^+}^4 \mu^{4 \epsilon} 
\left((g_-^{qZq})^4+(g_+^{qZq})^4\right)$, where
$N_C = 3$ is the number of colors in QCD, $Q_{e^+}$ is the electric charge 
of the positron, $4 - 2\epsilon$ is the number of space-time dimensions, 
and $g_\pm^{qZq}$ are defined by Eq.(3) of Ref.~\cite{Ohnemus:1991za}.
The definition of the function $F(t,u)$ is somewhat lengthy and can be
found in Appendix C of Ref.~\cite{Ohnemus-Owens} (cf. Eqs.~(C1) and (C2)).
The function ${\cal V}(t,u)$ depends on the kinematic
correlation between the initial and final states through its dependence 
on $t
$ and $u$. In the $m_Z \to 0$ limit, $F(t,u)+F(u,t)$ reduces to $%
F^{virt}(t,u)$ of the di-photon case which is given in Ref.~\cite
{Balazs-Berger-Mrenna-Yuan}.\footnote{%
This is connected to the fact that as $m_Z \to 0$
the virtual corrections of the $Z^0$ pair
and di-photon productions are the same (up to the couplings), which is
apparent when comparing Eq.(11) of Ref.~\cite{Bailey-Ohnemus-Owens} and 
Eq.(12) of Ref.~\cite{Ohnemus-Owens}, after including a missing factor 
of $1/(16 \pi s)$ in the latter equation.}

The non-perturbative function used in this study is \cite{Ladinsky-Yuan} 
\[
\widetilde{W}_{q\overline{q}}^{NP}(b,Q,Q_0,x_1,x_2)={\rm exp}\left[
-g_1b^2-g_2b^2\ln \left( {\frac Q{2Q_0}}\right) -g_1g_3b\ln {(100x_1x_2)}%
\right] , 
\]
with $g_1=0.11~{\rm GeV}^2$, $g_2=0.58~{\rm GeV}^2$, 
$g_3=-1.5~{\rm GeV}^{-1}$, and $Q_0=1.6~{\rm GeV}$. 
These values were
fit for the CTEQ2M parton distribution function, with the canonical
choice of the renormalization constants, i.e. $C_1=C_3=b_0$ and $C_2=1$,
and $b_{max}=0.5~{\rm GeV}^{-1}$ was used.  
In principle, these coefficients should be refit for CTEQ4M distributions 
\cite{CTEQ4} used in this study.
We have checked that using the updated fit in Ref.~\cite{Yuan_Moriond} 
does not change largely our conclusion because at the LHC and Tevatron
energies the perturbative Sudakov contribution is more important
compared to that in the low energy fixed target experiments.

Before concluding this section we note that for the di-photon
production, we use the formalism described in 
Ref.~\cite{Balazs-Berger-Mrenna-Yuan} 
to include the $gg\to \gamma \gamma X$ contribution, in which part of
the higher order corrections has been included via resummation. 
Since a gauge invariant calculation of the $gg\to Z^0 Z^0 g$ cross section in
the SM involves diagrams with the Higgs particle, 
we shall defer its discussion to a separate work ~\cite{Balazs-Yuan_Higgs}.

\section{Numerical Results}

We implemented our analytic results in the ResBos Monte Carlo event
generator \cite{Balazs-YuanPRD}. 
As an input we use the following electroweak parameters \cite{PDB}: 
\begin{eqnarray}
G_F=1.16639\times 10^{-5}~{\rm GeV}^{-2},~~m_Z=91.187~{\rm GeV},
~~m_W=80.41~{\rm GeV},~~\alpha (m_Z)=\frac 1{128.88}.
\nonumber
\end{eqnarray}
In the on-shell renormalization scheme we define the effective weak mixing 
angle
\begin{eqnarray}
\sin^2 \theta_w^{eff} = 1 - \frac{m_W^2}{\rho m_Z^2},
\nonumber
\end{eqnarray}
with
\begin{eqnarray}
\rho = \frac{m_W^2}{m_Z^2}
\left( 1 - \frac{\pi \alpha (m_Z)}{\sqrt{2} G_F m_W^2} \right)^{-1}.
\nonumber
\end{eqnarray}

In Eq.(\ref{Eq:DefgLgR}), the coupling of the $Z^0$ boson to fermions, $g$,
is defined using the improved Born approximation: 
\[
g^2=4\sqrt{2}G_F(c_w^{eff})^2m_Z^2\rho,
\]
with $c_w^{eff} = \sqrt{1 - \sin^2 \theta_w^{eff}}$, 
the cosine of the effective weak mixing angle.
(In Eq.~(\ref{Eq:DefgLgR}) $c_w$ is identified with $c_w^{eff}$.)
We use the NLO expression for the running strong and electroweak couplings
$\alpha _s(\mu)$ and $\alpha(\mu)$, as well as the NLO 
parton distribution function CTEQ4M (defined in the modified
minimal subtraction, i.e. ${\overline {\rm MS}}$, scheme), 
unless stated otherwise. 
Furthermore, in all cases we set the renormalization scale equal to the
factorization scale: $\mu_R=\mu_F=Q$.

Table~\ref{tbl:Total} summarizes the total rates for the leading order 
(LO), i.e. ${\cal O}(\alpha_s^0)$,
and resummed photon- and $Z^0$-pair production cross sections for the 
LHC and the Tevatron.
For the lowest order calculation we show results using LO (CTEQ4L) and 
NLO (CTEQ4M) parton distributions, 
because there is a noticeable difference
due to the PDF choice.
As it was discussed in Ref.~\cite{Balazs-YuanPRD}, 
the resummed total rate
is expected to reproduce the ${\cal O}(\alpha_s)$ rate, 
provided that in the resummed calculation the $A^{(1)}$, $B^{(1)}$ and 
$C^{(1)}$ coefficients and the ${\cal O}(\alpha_s)$ $Y$ piece are included, 
and the $Q_T$ distribution is described by the resummed result for 
$Q_T \leq Q$ and by the ${\cal O}(\alpha_s)$ result for $Q_T > Q$.
In our present calculation we added the $A^{(2)}$ coefficient to include
the most important higher order corrections in the Sudakov exponent.
Our matching prescription (cf. Ref.~\cite{Balazs-YuanPRD})
is to switch from the resummed prediction to the 
fixed-order perturbative calculation as they cross around $Q_T \sim Q$.
This switch is performed for any given $Q$ 
and $y$ of the photon or $Z^0$ boson pairs.
In the end, the total cross section predicted by our resummed 
calculation is about the same as that predicted by the NLO calculation. 
The small difference of those two predictions  
can be interpreted as an estimate of the contribution beyond
the NLO.

\subsection{$Z^0$ pair production at the LHC}

In the LHC experiments the $H \to Z^0 Z^0 $ channel can be identified 
through the decay products of the $Z^0$ bosons. The detailed experimental 
kinematic cuts for this process are given in Ref.~\cite{AtlasTechProposal}.
Since the aim of this work is not to analyze the decay kinematics of the 
background, rather to present the effects of the initial--state soft--gluon 
radiation, following Ref.~\cite{Ohnemus-Owens} 
for the LHC energies we restrict the rapidities of each $Z^0$ bosons as: 
$|y^Z|<3.0$. We do not apply any other kinematic cuts.
The total rates are given in Table~\ref{tbl:Total}.
Our ${\cal O}(\alpha_s^0)$ rates are in agreement with that of 
Ref.~\cite{Ohnemus-Owens} when calculated using the same PDF.
We expect the resummed rate to be higher than the 
${\cal O}(\alpha_s)$ rate due to the inclusion of the $A^{(2)}$ term.
Indeed, our $K$ factor, defined as the ratio of the resummed to the LO rate
using the same PDF in both calculations, is higher than the naive soft gluon
$K$-factor ($K_{\rm naive} = 1 + 8\pi \alpha_s(Q)/9 \sim 1.3$) of 
Ref.~\cite{Barger-Lopez-Putikka}, which estimates the NLO 
corrections to the production rate of $q\bar{q}\to Z^0 Z^0 X$ 
in the DIS (deep-inelastic scattering) scheme.
Our $K$-factor approaches the naive one with the increase of the 
center-of-mass energy, as expected.

The rates for the different subprocesses of the $Z^0$ boson pair production
are given in Table~\ref{tbl:SubTotalZZ}.
At the LHC the $q g \to Z^0 Z^0 X$ subprocess contributes
about 25\% of the $q q + q g \to Z^0 Z^0 X$ rate.
The $K$-factor is defined as the ratio $\sigma(q\bar{q}+qg\to Z^0 Z^0 X)
/\sigma(q\bar{q}\to Z^0 Z^0)$, which is about 1.4 for using CTEQ4M PDF.

Figs.~\ref{fig:ZZLHCQT}--\ref{fig:ZZLHCQypT} show our results for proton-proton  
collisions at the LHC energy, $\sqrt{S}=14$ GeV. 
Fig.~\ref{fig:ZZLHCQT} shows the transverse momentum distribution of $Z^0$ 
pairs. 
The NLO (${\cal O}(\alpha_s)$) prediction for the 
$q{\bar q} + q g \to Z^0 Z^0 X$ 
subprocesses, shown by the dotted curve, is singular as $Q_T \to 0$.
This singular behavior originates from the contribution of terms which
grow at least as fast as $1/Q_T^2$ or $ln(Q^2/Q_T^2)/Q_T^2$.
This, so-called asymptotic part, is shown by the dash-dotted curve,
which coincides with the $O(\alpha_s)$ distribution as $Q_T \to 0$.
After exponentiating these terms, the distribution
is well behaved in the low $Q_T$ region, as shown by the solid curve. 
The resummed curve matches the $O(\alpha_s)$ curve at about 
$Q_T=320$ GeV.   
Following our matching prescription described in the previous section, 
we find that this matching takes place around $Q_T = 300$ GeV, 
depending on the actual values of $Q$ and $y$.
Fig.~\ref{fig:ZZLHCQT} also shows that at the LHC there
is a substantial contribution from $q g$ scattering,
which is evident from the difference between the solid and 
dashed curves, where the dashed curve is the resummed contribution from 
the $q \bar q \to Z^0 Z^0 X$ subprocesses.

In Fig.~\ref{fig:ZZLHCInt} we give the integrated distributions, defined as
\begin{equation}
\frac{d\sigma}{dQ_T^{\min }} =
\int_{Q_T^{\min }}^{Q_T^{\max}}dQ_T\;
\frac{d\sigma}{dQ_T}, 
\label{eq:Integrated}
\end{equation}
where $Q_T^{\max}$ is the largest $Q_T$ allowed by the phase space. 
In the NLO calculation, this distribution grows
without bound near $Q_T^{\min }=0$, as a result of the singular behavior 
of the scattering amplitude when $Q_T \rightarrow 0$. 
It is clearly shown by Fig.~\ref{fig:ZZLHCInt} that the $Q_T$ 
distribution of the resummed calculation is different from that of 
the NLO calculation.
The different shapes of the two curves in Fig.~\ref{fig:ZZLHCInt}
indicates that the predicted $Z^0$ pair production rates, with a minimal value
of the transverse momentum $Q_T$, are different in the two calculations. This
is important at the determination of the background for the detection of a
Higgs boson even with moderately large transverse momentum.
For $Q_T^{min}=50$ GeV, the resummed cross section is 
about 1.5 times of the NLO cross section.

The invariant mass and the rapidity distributions of the $Z^0$ boson pairs, 
and the transverse momentum distribution of the individual $Z^0$ bosons
are shown in Fig~\ref{fig:ZZLHCQypT}.
When calculating the production rate as the function of the $Z^0$ pair 
invariant mass,
we integrate the $Q_T$ distribution for any $Q$, and $y$. 
When plotting the transverse momentum distributions of the individual 
$Z^0$ bosons, we include both of the $Z^0$ bosons per event.
 In the shape of the invariant mass and 
rapidity distributions we do not expect large
deviations from the NLO results. 
Indeed, the shape of our invariant mass distribution agrees with that in 
Ref.~\cite{Ohnemus-Owens}.
However, the resummed transverse momentum distribution 
$P_T^Z$ of the individual $Z^0$ bosons
is slightly broader than the NLO distribution 
(not shown in Fig~\ref{fig:ZZLHCQypT}, cf. Ref.~\cite{Ohnemus-Owens}).
This is expected because, in contrast with the NLO distribution, the resummed
transverse momentum distribution of the $Z^0$ boson pair is finite as 
$Q_T \rightarrow 0$ so that the $P_T^Z$ 
distribution is less peaked.

In Figs.~\ref{fig:ZZLHCQTCi} and \ref{fig:ZZLHCQTgi} we show the
dependence of the resummed result on the values of the renormalization
constants $C_i$ ($i=1,2$), and the values of the non-perturbative
parameters $g_i$ ($i=1,2,3$), respectively. As Eq.~(\ref{Eq:WTwi})
shows, both the Sudakov exponent ${\cal S}_{ij}$ and the Wilson
coefficients ${\cal C}_{i/h}$ depend on the renormalization constants
$C_1$ and $C_2$. The scale $C_1/b$ determines the onset of the
non-perturbative physics, and $C_2 Q$ specifies the scale of the hard
scattering process. We vary both $C_1$ and $C_2$ by a factor of 2. In
Fig.~\ref{fig:ZZLHCQTCi}, we show that the resummed calculation using
the canonical $C_1$ and $C_2$ values (upper solid curve) hardly differs
from the one which uses $C_1 = b_0/2$ and $C_2 = 1/2$ (lower solid
curve). This difference is certainly smaller than the difference between
the resummed and fixed order (dashed) curves in the $Q_T^{ZZ}=50-100$
GeV region. In Fig.~\ref{fig:ZZLHCQTgi}, we show two resummed curves,
one with the non-perturbative parameters given at the end of Section
II.B (solid curve), and one with the following non-perturbative
parameters: $g_1=0.15~{\rm GeV}^2$, $g_2=0.48~{\rm GeV}^2$,
$g_3=-.58~{\rm GeV}^{-1}$, and $Q_0=1.6~{\rm GeV}$ (dashed curve, c.f.
Ref~\cite{Yuan_Moriond}). There is some difference only in the lowest
$Q_T^{ZZ}$ region ($Q_T^{ZZ}<30$ GeV), and this difference is negligible
compared to the difference between the resummed and NLO (dotted)
calculations. Based on these results, we conclude that the CSS
resummation gives a stable prediction for the gauge boson pair
production at the LHC energy, and the same conclusions also hold for the
Tevatron.

\subsection{$Z^0$ Pair Production at the upgraded Tevatron}

After the upgrade of the Fermilab Tevatron, there are
more than $3\times 10^3$ $Z^0$ boson pairs to be produced.
Since this data sample can be used to test
the tri-linear gauge boson couplings \cite{www}, we also give our 
results for  
the upgraded Tevatron with proton--anti-proton collisions
at a center-of-mass energy of $2$ TeV. 
Our kinematic cuts constrain the rapidity of both of the $Z^0$
bosons such that $|y^Z| < 3$.
Both the LO and resummed total rates are listed in Table~\ref{tbl:Total}.
The ratio $\sigma(q\bar{q}+qg\to Z^0 Z^0 X)/\sigma(q\bar{q}\to Z^0 Z^0)$ 
is about 1.6, which is larger than the naive soft gluon $K$-factor of 
1.3. 
Table~\ref{tbl:SubTotalZZ} shows that  
$q g \to Z^0 Z^0 X$ partonic subprocess contributes only a 
small amount (about 3\%) at this energy, in contrast to
25\% at the LHC.

Figs.~\ref{fig:ZZTevQypT}--\ref{fig:ZZTevInt} show the resummed
predictions for the upgraded Tevatron. 
The invariant mass and rapidity distributions of $Z^0$ boson pairs, 
and the transverse momentum distribution of the individual $Z^0$ bosons
are shown in Fig.~\ref{fig:ZZTevQypT}. 
The solid curve shows the resummed contributions from the 
$q {\bar q} + q g \to Z^0 Z^0 X$ subprocess. 
The resummed contribution from the $q {\bar q} \to Z^0 Z^0 X$ subprocess
is shown by the dashed curve.
The leading order $q {\bar q} \to Z^0 Z^0$ cross section is also 
shown, by the dash-dotted curve.
The invariant mass distribution of the $q {\bar q} + q g \to Z^0 Z^0 X$
subprocess is in agreement with the NLO result of Ref.~\cite
{Ohnemus-Owens}, when calculated for $\sqrt{S}=1.8$ TeV.
From this figure we also find that the
contribution from the $q g \to Z^0 Z^0 X$ subprocess at the energy of the 
Tevatron is very small.

In Fig.~\ref{fig:ZZTevQT} we compare the NLO and resummed 
 distributions of the transverse momentum of the $Z^0$ pair.
The figure is qualitatively similar to that at the LHC, 
as shown in Fig.~\ref{fig:ZZLHCQT}. 
The resummed and the NLO curves merge at about 100 GeV.
The resummed contribution from the $q {\bar q} \to Z^0 Z^0 X$
subprocess is shown by the dashed curve, which clearly 
dominates the total rate.

In Fig.~\ref{fig:ZZTevInt} we show the integrated distribution 
$d\sigma/dQ_T^{min}$ for $Z^0$ boson pair production at the 
upgraded Tevatron. 
The figure is qualitatively the same as that for the LHC 
(cf. Fig.~\ref{fig:ZZLHCInt}).
The NLO curve runs well under the resummed one in the $Q_T^{\min }<$ 
80 GeV region, and the $Q_T$ distributions from the NLO and the resummed 
calculations have different shapes even in the region where $Q_T$ is of the 
order 60 GeV. 
For $Q_T^{min}=30$ GeV, the resummed rate is about 1.5 times of the NLO rate.

\subsection{Di-photon production at the LHC}

Photon pairs from the decay process $H \to \gamma \gamma $ can
be directly detected at the LHC. 
When calculating its most important background rates, 
we impose the kinematic cuts on the final state photons
that reflect the optimal detection capabilities 
of the ATLAS detector \cite{AtlasTechProposal}:

$p_T^\gamma >25$ GeV, for the transverse momentum of each photons,

$\left| y^\gamma \right| <2.5$, for the rapidity of each photons, and

$p_T^1/(p_T^1+p_T^2)<0.7$, to suppress the fragmentation contribution,
where $p_T^1$ is the transverse momentum of the photon 
with the higher $p_T$ value.\\
We also apply a $\Delta R=0.4$ separation cut on the photons, but our results
are not sensitive to this cut. 
(This conclusion is similar to that in 
Ref.~\cite{Balazs-Berger-Mrenna-Yuan}.)
The total rates and cross sections from the different partonic subprocesses 
are presented in Tables~\ref{tbl:Total} and ~\ref{tbl:SubTotalAA}.
We have incorporated part of the higher order contributions to this process
by including $A^{(2)}$ in the Sudakov factor and $C^{(1)}_{gg}$ in 
the Wilson coefficient functions (cf. Ref.~\cite{Balazs-Berger-Mrenna-Yuan}).
Within this ansatz, up to ${\cal O}(\alpha_s^3)$, the 
$gg\to \gamma \gamma X$ rate is about 24 pb, which increases the total 
$K$-factor by almost 1.0.
The leading order $gg\to \gamma \gamma$ rate, via the box diagram, is about 
22 pb and 14 pb for using the LO PDF CTEQ4L and the NLO PDF CTEQ4M, 
respectively. The large difference mainly due to the differences in the
strong coupling constants used in the two calculations%
\footnote{When using the CTEQ4L PDF, we consistently use the LO running 
coupling constant $\alpha_s$.}:
CTEQ4L requires $\alpha_s(m_Z)=0.132$, while for CTEQ4M $\alpha_s(m_Z)=0.116$.
The ratio 
$\sigma(q\bar{q}+qg\to \gamma \gamma X)/\sigma(q\bar{q}\to \gamma \gamma)$ 
is 1.5, and 
$\sigma(gg\to \gamma \gamma X)/\sigma(q\bar{q}\to \gamma \gamma)$ 
is about 1. 
Hence, the ratio of the resummed and the ${\cal O}(\alpha_s^0)$ rates, 
is quite substantial.

Figs.~\ref{fig:AALHCQypT}--\ref{fig:AALHCInt} show our predictions
for distributions of di-photons produced at the LHC. 
In Fig.~\ref{fig:AALHCQypT} we plot
the invariant mass and rapidity distribution of the photon pairs, 
and the transverse momentum distribution of the individual photons.
When plotting the transverse momentum distributions of the 
individual photons
we include both photons per event.
The total (upper solid) and the resummed 
$q{\bar q} + q g \to \gamma \gamma X$ (dashed), 
$q{\bar q}\to \gamma \gamma X$ (dotted), 
$g g \to \gamma \gamma X$ (dash-dotted), 
and the fragmentation (lower solid), 
as well as the leading order
 $q{\bar q} \to \gamma \gamma$ (middle solid)
contributions are shown separately.
The ratio of the resummed and the LO distributions is about 2.5 
which is consistent with the result in Table~\ref{tbl:Total}. 
The relative values of the contributions from each subprocesses reflect 
the summary given in Table~\ref{tbl:SubTotalAA}.

Fig.~\ref{fig:AALHCQT} shows various
contributions to the transverse momentum of the photon pair.
At low $Q_T$ values ($Q_T \ll Q$), the $q{\bar q} \to \gamma \gamma X$ 
contribution is larger than the
$qg \to \gamma \gamma X$ contribution,
while at high $Q_T$ values ($Q_T>Q$), the 
$qg \to \gamma \gamma X$ subprocess becomes more important.
The $gg$ contribution dominates the total rate in low $Q_T$ region, and
the kink in the $gg$ curve at about 50 GeV indicates the need for the 
inclusion of the complete ${\cal O}(\alpha_s^3)$ $gg\to\gamma\gamma g$ 
contribution.
(Recall that our prediction for the $gg$ contribution 
at ${\cal O}(\alpha_s^3)$ only holds for small $Q_T$, where the 
effect of the initial-state soft-gluon radiation is relatively
more important for a fixed $Q$.)
 
In Fig.~\ref{fig:AALHCInt} we give the integrated cross section 
as the function of the transverse momentum of the 
photon pair produced at the LHC. 
Similarly to the $Z^0$ pair production, 
there is a significant shape difference 
between the resummed and the NLO curves in the low to mid $Q_T$ region.
For $Q_T^{min}=50$ GeV, the resummed rate is about 1.5 times of the NLO
rate.

\subsection{Di-photon production at the upgraded Tevatron}

In Ref.~\cite{Balazs-Berger-Mrenna-Yuan}, we have presented the 
predictions of the CSS resummation formalism for the di-photon production
at the Tevatron with $\sqrt{S}=1.8$ TeV, and compared with the data 
\cite{CDFDiPhoton,D0DiPhoton}.
In this paper, we show the results for the upgraded Tevatron with 
$\sqrt{S}=2.0$ TeV.
We use the same kinematic cuts which were used in 
Ref.~\cite{Balazs-Berger-Mrenna-Yuan}:

$p_T^\gamma > 12$ GeV, for the transverse momentum of each photons,

$\left| y^\gamma \right| < 0.9$, for the rapidity of each photons. \\
An isolation cut of $\Delta R=0.7$ is also applied.
The total cross sections and the rates of the different subprocesses 
are given by Tables~\ref{tbl:Total} and ~\ref{tbl:SubTotalAA}.
The ratio of the $q\bar{q}+qg\to \gamma \gamma X$ and 
$q\bar{q}\to \gamma \gamma$ rates is about 1.5, similar to
that at the LHC.
The leading order rate for the $gg\to \gamma \gamma$ subprocess is
about 6.0 pb and 4.3 pb for using CTEQ4L and CTEQ4M PDF, respectively.
The NLO rate for $gg\to \gamma \gamma g$ is estimated to be 8.3 pb, using 
the approximation described in Ref.~\cite{Balazs-Berger-Mrenna-Yuan}, which 
is about the same magnitude 
as the leading order $q\bar{q} \to \gamma \gamma$ rate.
From our estimate of the NLO $gg$ rate, we expect that the 
complete ${\cal O}(\alpha_s^3)$ contribution will be important
for photon pair production at the Tevatron.

Figs.~\ref{fig:AATevQypT}--\ref{fig:AATevInt} show our results
for photon pairs produced at the upgraded Tevatron. 
The resummed predictions for the 
invariant mass and rapidity distributions of the photon pairs, and the
transverse momentum distribution of the individual photons are 
shown in Fig.~\ref{fig:AATevQypT}.
In Fig.~\ref{fig:AATevQT} we also plot
the contributions to the transverse momentum of 
the photon pair from the 
$q{\bar q} + q g \to \gamma \gamma X$ (dashed), 
$q{\bar q}\to \gamma \gamma X$ (dotted), 
$gg\to\gamma\gamma g$ (dash-dotted), 
and the fragmentation (lower solid) subprocesses, separately. 
The leading order $q{\bar q}\to\gamma\gamma$ 
cross section (middle solid) is also plotted.
In the low $Q_T$ region, the $gg$ and the $q\bar{q}$ rates
are about the same, and the $qg$ rate becomes more important in the large
$Q_T$ region. 
Furthermore, after imposing the above kinematic cuts,
the fragmentation contribution is found to be unimportant.

Fig.~\ref{fig:AATevInt} shows the integrated $Q_T$ distribution. 
The qualitative features of these distributions are the same as 
those predicted for the LHC.
For $Q_T^{min}=10$ GeV, the resummed cross section is 
about twice of the NLO cross section.

\section{Conclusions}

In this work we studied the effects of the initial--state multiple 
soft--gluon emission on the total rates and various distributions of the 
most important background processes 
($pp, p{\bar p} \to \gamma \gamma X, Z^0Z^0 X$)
to the detection of the Higgs boson at the LHC.
We applied the extended CSS formalism to resum the large
logarithms induced by the soft--gluon radiation.
We found that for the $q {\bar q}$ and $q g$ initiated processes, 
the total cross sections and the invariant mass distributions
of the photon and $Z^0$ boson pairs are in agreement with the fixed
order calculations.
From our estimate of the NLO rate of the $gg$ initiated 
process, we expect that the 
complete ${\cal O}(\alpha_s^3)$ contribution will be important
for photon pair production at the Tevatron.
We showed that the resummed and the NLO transverse momentum distributions
of the $Z^0$ and photon pairs are substantially different for 
$Q_T\lesim Q/2$.
In terms of the integrated cross section above a given $Q_T^{\rm min}$,
this difference can be as large as 50\% in the low to mid-range of 
$Q_T^{\rm min}$.
Using the resummation calculation, we are able to give
a reliable prediction of the $Q_T$ and any other distribution 
in the full kinematical region at the LHC and the Tevatron, 
even in the presence of kinematic cuts.
Since the bulk of the signal is in the low transverse momentum region, 
we conclude that the difference between the NLO and resummed predictions 
of the background rates will be essential when extracting the signal of 
the Higgs boson at hadron colliders.

\section*{Acknowledgments}

We thank the CTEQ collaboration, E. Berger, S. Mrenna, W. Repko and 
C. Schmidt for many invaluable discussions, J. Huston and M. Abolins 
for help with the ATLAS parameters and for useful conversations.
C.B. also thanks the Theory Division of FermiLab for 
the invitation and hospitality.
This work was supported in part by the NSF under grant PHY--9802564.

\vspace{0.4cm} 


\newpage

\TblTotal
\TblSubTotalZZ
\TblSubTotalAA

\clearpage

\FigDiagrams
\FigZZLHCQT
\FigZZLHCInt
\FigZZLHCQypT
\FigZZLHCQTCi
\FigZZLHCQTgi
\FigZZTevQypT
\FigZZTevQT
\FigZZTevInt
\FigAALHCQypT
\FigAALHCQT
\FigAALHCInt
\FigAATevQypT
\FigAATevQT
\FigAATevInt

\end{document}